\documentclass[12pt,onecolumn,draftclsnofoot,letterpaper]{IEEEtran}
\usepackage{amsmath,amsthm,amsfonts,amssymb,bm,mathrsfs}
\usepackage{graphicx,subfigure}
\usepackage[bookmarks=false,colorlinks,citecolor=black, filecolor=black, linkcolor=black,urlcolor=black]{hyperref}
\usepackage{cite}
\usepackage{url}

\usepackage{algorithm}
\usepackage{algorithmic}
\usepackage{multirow,booktabs,threeparttable}
\graphicspath{{DoubleColumns/}}
\theoremstyle{remark}

\newtheorem{lemma}{Lemma}
\newtheorem{definition}{Definition}
\newtheorem{theorem}{Theorem}

\begin{document}
%
\title{TACT:  A Transfer Actor-Critic Learning Framework for Energy Saving in Cellular Radio Access Networks}


\author{\IEEEauthorblockN{Rongpeng Li{\IEEEauthorrefmark{1}},  Zhifeng Zhao{\IEEEauthorrefmark{1}}, Xianfu Chen{\IEEEauthorrefmark{2}}, Jacques Palicot{\IEEEauthorrefmark{3}} and Honggang Zhang{\IEEEauthorrefmark{1}\IEEEauthorrefmark{4}}}\\
\IEEEauthorblockA{\IEEEauthorrefmark{1} Zhejiang University, Zheda Road 38, Hangzhou 310027, China\\
Email: \{lirongpeng, zhaozf, honggangzhang\}@zju.edu.cn\\}
\IEEEauthorblockA{\IEEEauthorrefmark{2}VTT Technical Research Centre of Finland, P.O. Box 1100, FI-90571 Oulu, Finland\\
Email: xianfu.chen@vtt.fi\\}
\IEEEauthorblockA{\IEEEauthorrefmark{3}Sup\'elec, Avenue de la Boulaie CS 47601, Cesson-S\'evign\'e Cedex, France\\
Email: jacques.palicot@supelec.fr\\}
\IEEEauthorblockA{\IEEEauthorrefmark{4}Universit\'e Europ\'eenne de Bretagne (UEB) \& Sup\'elec, Avenue de la Boulaie CS 47601, Cesson-S\'evign\'e Cedex, France}}

\maketitle

\begin{abstract}
Recent works have validated the possibility of improving energy efficiency in radio access networks (RANs), achieved by dynamically turning on/off some base stations (BSs). In this paper, we extend the research over BS switching operations, which should match up with traffic load variations. Instead of depending on the dynamic traffic loads which are still quite challenging to precisely forecast, we firstly formulate the traffic variations as a Markov decision process. Afterwards, in order to foresightedly minimize the energy consumption of RANs, we design a reinforcement learning framework based BS switching operation scheme. Furthermore, to speed up the ongoing learning process, a transfer actor-critic algorithm (TACT), which utilizes the transferred learning expertise in historical periods or neighboring regions, is proposed and provably converges. In the end, we evaluate our proposed scheme by extensive simulations under various practical configurations and show that the proposed TACT algorithm contributes to a performance jumpstart and demonstrates the feasibility of significant energy efficiency improvement at the expense of tolerable delay performance. 
\end{abstract}



\section{Introduction}
The explosive popularity of smartphones and tablets has ignited a surging traffic load demand for radio access and has been incurring massive energy consumption and huge greenhouse gas emission \cite{wu_green_2012,zhang_energy_2010}. Specifically speaking, the information and communication technology (ICT) industry accounts for 2\% to 10\% of the world's overall power consumption \cite{marsan_optimal_2009} and has emerged as one of the major contributors to the world-wide $\text{CO}_2$ emission. Besides that, there also exist economical pressures for cellular network operators to reduce the power consumption of their networks. It's envisioned that the electricity bill will doubly enlarge in five years for China Mobile \cite{china_mobile_research_institute_c-ran:_2010}. Meanwhile, the energy expenditure accounts for a significant proportion of the overall cost. Therefore, it's quite essential to improve the energy efficiency of ICT industry.

Currently, over $80\%$ of the power consumption takes place in the radio access networks (RANs), especially the base stations (BSs) \cite{fettweis_ict_2008}. The reason behind this is largely due to that the present BS deployment is on the basis of peak traffic loads and generally stays active irrespective of the heavily dynamic traffic load variations \cite{son_base_2011,peng_traffic-driven_2011}. Recently, there has been a substantial body of work towards traffic load-aware BS adaptation \cite{niu_tango:_2011} and the authors have validated the possibility of improving energy efficiency from different perspectives. Luca Chiaraviglio \textit{et al.} \cite{chiaraviglio_energy-aware_2008} showed the possibility of energy saving by simulations. \cite{niu_cell_2010} and \cite{li_gm-pab:_2012} proposed how to dynamically adjust the working status of BS, depending on the predicted traffic loads. However, to reliably predict the traffic loads is still quite challenging, which makes these works suffering in practical applications. On the other hand, \cite{oh_energy_2010} and \cite{zhou_green_2009} presented dynamic BS switching algorithms with the traffic loads a prior and preliminarily proved the effectiveness of energy saving.

Besides, it is also found that turning on/off some of the BSs will immediately affect the associated BS of a mobile terminal (MT). Moreover, subsequent choices of user associations in turn lead to the traffic load differences of BSs. Hence, any two consecutive BS switching operations are correlated with each other and current BS switching operation will also further influence the overall energy consumption in the long run. In other words, the expected energy saving scheme must be \textit{foresighted} while minimizing the energy consumption. It should concern its effect on both the current and future system performance to deliver a visionary BS switching operation solution.

\begin{figure}
\centering
\includegraphics[width=0.75\textwidth]{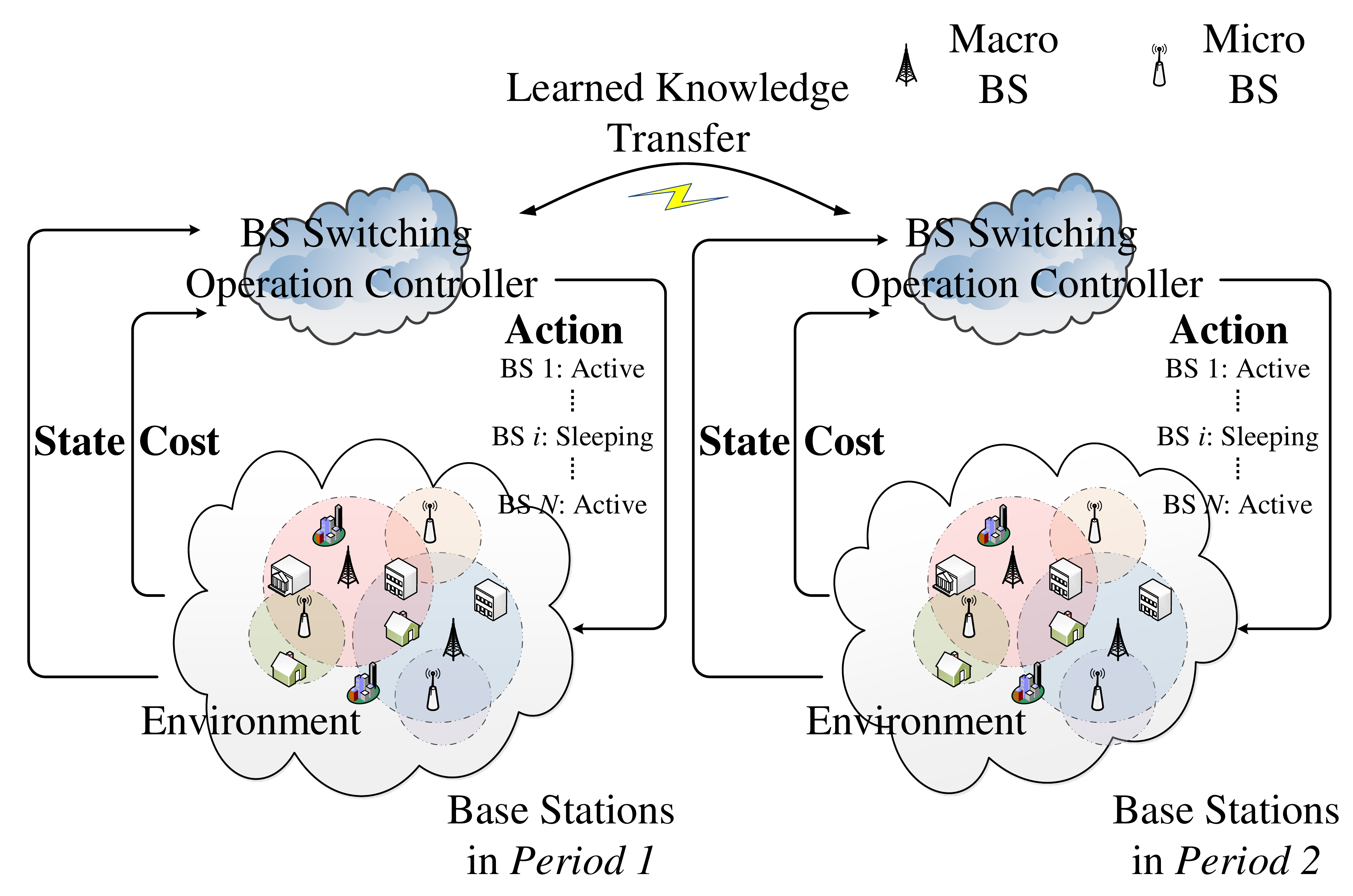}
\caption{Transfer learning for reinforcement learning in BS switching operation scenario.}
\label{fig:BSDeploymentMotivation}
\end{figure}

The authors in \cite{son_base_2011} presented a partially foresighted energy saving scheme which combines BS switching operation and user association by giving a heuristic solution on the basis of a stationary traffic load profile. In this paper, we try to solve this problem from a different perspective. Instead of predicting the volume of traffic loads, we apply a Markov decision process (MDP) to model the traffic load variations. Afterwards, the solution to the formulated MDP problem can be attained by making use of actor-critic algorithm \cite{konda_actor-critic--type_1999,konda_actor-critic_2000}, a reinforcement learning (RL) approach \cite{sutton_reinforcement_1998}, one advantage of which is that there is no necessity to possess a prior knowledge about the traffic loads within the BSs. On the other hand, given the centralized structure of cellular networks, energy saving will significantly benefit from a literally existing centralized BS switching operation controller such as the base station controller (BSC) in second generation (2G) cellular networks or the radio network controller (RNC) in third generation (3G) or long term evolution (LTE) cellular networks rather than a distributed one. As a result, we assume that a BS switching operation controller exists within the reinforcement learning framework, as illustrated in Fig. \ref{fig:BSDeploymentMotivation}. The controller would firstly estimate the traffic load variations based on the on-line experience. Afterwards, it can select one of the possible BS switching operations under the estimated circumstance and then decreases or increases the probability of the same action to be later selected on the basis of the required cost. Here, the cost primarily focuses on the energy consumption due to such a BS switching operation and also takes the performance metric into account to ensure the user experience. After repeating the actions and knowing the corresponding costs, the controller would know how to switch the BSs for one specific traffic load profile. Moreover, with the MDP model, the resulting BS switching strategy is foresighted, which would improve energy efficiency in the long run. 

However,  it usually take some time for the RL approaches to be convergent to the optimal solution in terms of the whole cost \cite{berenji_convergent_2003,woergoetter_reinforcement_2008}. Hence, the direct application of the RL algorithms may sometimes get into trouble, especially for a scenario where a BS switching operation controller usually takes charge of tens or even hundreds of BSs \cite{li_gm-pab:_2012}. Fortunately, the periodicity and mobility of human behavior patterns make the traffic loads exhibit some temporal and spatial relevancies \cite{zhou_predictability_2012}, thus making the traffic load-aware BS switching strategies at different moments or neighboring regions relevant. Therefore, we could deal with the application issue by utilizing the conceptual idea of transfer learning (TL) \cite{taylor_transfer_2009}. TL, which mostly concerns how to recognize and apply the knowledge learned from one or more previous tasks (\textit{source tasks}) to more effectively learn to solve a novel but related task (\textit{target task}) \cite{aha_case-based_2009},  is intuitively appealing, cognitively inspired, and has led to a burst of research activities \cite{taylor_transfer_2009,pan_survey_2010,aha_case-based_2009,celiberto_luiz_a._using_2010}. By transferring the learned BS switching operation strategy at historical moments or neighboring regions (source tasks), TL could exploit the temporal and spatial relevancy in the traffic loads and speed up the on-going learning process in regions of interest (target tasks) as depicted in Fig \ref{fig:BSDeploymentMotivation}.  As a result, the learning framework of BS switching operation is further enhanced by incorporating the idea of TL into the classical actor-critic algorithm (AC), namely the Transfer Actor-CriTic algorithm (TACT)  in this paper.

In a nutshell, our work proposes a reinforcement learning framework for energy saving in RANs. Compared to the previous works, this paper provides the following three key insights:
\begin{itemize}
\item Firstly, we show that the learning framework is feasible to save the energy consumption in RANs without the knowledge of traffic loads a prior. Moreover, the performance of the learning framework approaches that of the state-of-the-art scheme (SOTA) \cite{son_base_2011}, which is assumed to have fully knowledge of traffic loads. These preliminary results have already been presented in \cite{li_energy_2012}.
\item Secondly, we extend the idea of TL to the conventional RL algorithms and show that the proposed TACT algorithm outperforms the classical AC algorithm with a performance jumpstart.
\item Thirdly, this paper details the convergence analysis of the TACT algorithm and thereby contributes to the general literature in RL field, especially the general AC algorithm.
\end{itemize}

The remainder of the paper is organized as follows. In Section \ref{sec:problem}, we introduce the system model and formulate the traffic variation as an MDP. In Section \ref{sec:learning}, we talk about the energy saving scheme by the conventional RL framework. Section \ref{sec:tact} focuses on the incorporation of idea of TL into the conventional RL framework and investigates the convergence proof of the TACT algorithm. Section \ref{sec:analysis} evaluates the proposed schemes and presents the validity and effectiveness. Finally, we concludes this paper and presents several remaining works in Section \ref{sec:conclusion}.
\begin{table}
\centering
\caption{A list of the main symbols in the paper.}
\label{tb:notations}
\begin{tabular}{r | l}
\toprule
Symbol & Meaning\\
\midrule
$M=<\mathcal{S},\mathcal{A},\mathcal{P},C>$ & MDP Tuple: State Space $\mathcal{S}$, Action Space $\mathcal{A}$, \\ 
$\bm{s}^{(k)} \in \mathcal{S},\bm{a}^{(k)} \in \mathcal{A}$& State Transition Probability Function $\mathcal{P}$, \\
& and Cost Function $C$\\
superscript $(k)$ & Stage Number \\
$V^{\pi}(\bm{s})$ & Value Function $V$ w.r.t. Strategy $\pi$ and State $\bm{s}$ \\
$p(\bm{s},\bm{a})$ & Policy: Tendency to Select Action $\bm{a}$ under State $\bm{s}$ \\
$p_{\text{o}}$, $p_{\text{n}}$ and $p_{\text{e}}$ & Subscript o, n, e: Overall, Native and Exotic Policy\\
 $\delta(\bm{s}^{(k)},\bm{a}^{(k)})$  & TD Error under State $\bm{s}^{(k)}$ and Action $\bm{a}^{(k)}$\\
 \midrule
$\nu_1(\bm{s}^{(k)},k)$ & Occurrence of State $\bm{s}^{(k)}$ in the Previous $k$ Stages\\
$\hat{k}=\nu_2(\bm{s}^{(k)},\bm{a}^{(k)},k)$ & Occurrence of  $(\bm{s}^{(k)},\bm{a}^{(k)})$ in the Previous $k$ Stages\\
$\hat{p}_{\text{o}}(\hat{k})$ & Discrete Sequence: Evolution of $p_{\text{o}}^{(k)}(\bm{s},\bm{a})$\\
$\hat{p}^{(0)}(t)$ & Continuous Sequence: Interpolation Result of $\hat{p}_{\text{o}}(\hat{k})$\\
$\hat{p}^{(\hat{k})}(t)$ & Temporal Shifted Version of $\hat{p}^{(0)}(t)$\\
$\dot{\pi}(t)$, $\dot{V}(t)$ and $\dot{p}_{\text{o}}(t)$ & Derivative of $\pi(t)$, $V(t)$ and $p_{\text{o}}(\bm{s},\bm{a})$\\
$\alpha(\cdot)$, $\beta(\cdot)$, and $\zeta(\cdot)$ & Positive Step-Size Parameter in Learning Algorithms\\
$\lambda(x),1/\mu(x)$ & Arrival Rate  and File Size at Location $x$\\
$q_i$ & Constant Power Consumption Percentage for BS $i$\\
$\tau$ & Temperature: Positive Parameter\\
$\varsigma$ & Delay Performance Importance: Positive Parameter\\
\bottomrule
\end{tabular}
\end{table}
\section{System model and problem formulation}
\label{sec:problem}
\subsection{System model}
Beforehand, Table \ref{tb:notations} summarizes the most used notations in this paper.  

An RAN usually consists of multiple BSs while the traffic loads of BSs are usually fluctuating, thus often making BSs under-utilization. In this paper, we assume that there exists a region $\mathcal{L} \in \mathbb{R}^2$ served by a set of overlapped BSs $\mathcal{B}=\{1,\ldots,N\}$ as Fig. \ref{fig:BSDeploymentMotivation} depicts. In addition, we assume there exists a BS switching operation controller, which can timely know the traffic loads in these BSs at current stage and correspondingly determine the energy efficient working status of any BS (i.e., active/sleeping mode) at next stage in a centralized way. Beyond that, the paper focuses on downlink communication, i.e., from BSs to MTs. Meanwhile, the file transmission requests at a location $x \in \mathcal{L}$ arrive following a Poisson point process with arrival rate per unit area $\lambda(x)$ and file size $\frac{1}{\mu(x)}$ \cite{kim_alpha-optimal_2010,das_dynamic_2003,sang_coordinated_2008}. After that, the \textit{traffic load density} at a location $x \in \mathcal{L}$ is defined as $\lambda(x)/\mu(x)<\infty$ \cite{son_base_2011,kim_alpha-optimal_2010}. Therefore, the traffic load density can capture different spatial traffic variations. For example, a hotspot can be characterized by higher arrival rate or larger file size. Hence, when the set of BSs $\mathcal{B}_{\text{on}}$ is turned on, the traffic loads severed by BS $i \in \mathcal{B}_{\text{on}}$ can be represented as $\Gamma_i=\int_{\mathcal{L}}  I_i(x,\mathcal{B}_{\text{on}}) \lambda(x)/\mu(x)  \, \text{d}x$, whereas $I_i(x,\mathcal{B}_{\text{on}})=1$ is a user association indicator and denotes location $x$ is served by BS $i \in \mathcal{B}_{\text{on}}$ and vice versa. Otherwise, if a BS $i$ is in sleeping mode, i.e.,  $i\in \mathcal{B}\setminus\mathcal{B}_{\text{on}}$, the traffic loads are defined as zero, namely $\Gamma_i=0$. To demonstrate the temporal traffic load variations within one BS' coverage, i.e., $\mathcal{P}(\Gamma_i^{(k+1)}|\Gamma_i^{(k)})$ within the coverage of BS $i$, we partition the traffic loads $\Gamma_i$ into several segments and use a finite state indicator $s_i \in \mathcal{S}_i$ to describe one segment. Subsequently,  for the whole region of interest, a state vector $\bm{s}=\{s_1,\cdots,s_N\} \in \mathcal{S}=\mathcal{S}_1 \times \cdots \times \mathcal{S}_N$ is constructed to model the traffic load variations and constitutes a finite state Markov chain (FSMC).

Let's denote the transmission rate of a user located at $x$ and served by BS $i\in \mathcal{B}_{\text{on}}$ as $c_i(x,\mathcal{B}_{\text{on}})$. For analytical convenience, assume that $c_i(x,\mathcal{B}_{\text{on}})$ does not change over time, i.e., we do not consider fast fading or dynamic inter-cell interference. Instead, $c_i(x,\mathcal{B}_{\text{on}})$ is assumed as a time-averaged transmission rate in this paper, based on the fact that the time scale of user association is commonly much larger than the time scale of fast fading or dynamic inter-cell interference. Hence, the inter-cell interference is considered as static Gaussian-like noise, which is feasible under interference randomization or fractional frequency reuse \cite{son_base_2011,sang_coordinated_2008}. Beyond that, though $c_i(x,\mathcal{B}_{\text{on}})$ is location-dependent, it is also affected by the shadowing effect and thus not necessarily determined by the distance from the BS $i$.

Furthermore, the \textit{system load density} can be defined as the fraction of time required to deliver traffic loads from BS $i \in \mathcal{B}_{\text{on}}$ to location $x$, namely $\varrho_i(x)=\lambda(x)/\left(\mu(x)c_i(x,\mathcal{B}_{\text{on}}) \right)$. Analogous to the definition of traffic loads, the system loads for an active BS $i \in \mathcal{B}_{\text{on}}$ can be represented as $\rho_i=\int_{\mathcal{L}} \varrho_i(x) I_i(x,\mathcal{B}_{\text{on}}) \, \text{d}x$. Meanwhile, the system loads for a sleeping BS $i\in \mathcal{B}\setminus \mathcal{B}_{\text{on}}$ is defined as zero. Hence, the indicator set $\bm{I}=\{I_i(x,\mathcal{B}_{\text{on}})|i\in \mathcal{B}, x\in \mathcal{L}\}$ is feasible \cite{kim_alpha-optimal_2010} if each BS $i\in \mathcal{B}$ can serve $\rho_i<1$. Eventually, our goal is to choose certain active BSs and find a feasible user association indicator set to minimize the total cost. By exploiting the proposed learning framework, the controller can know the BS switching operation strategy at last without the prior knowledge of traffic loads. We will give the details in Section \ref{sec:learning}.
\subsection{Problem formulation}
In this paper, we primarily aim to minimize the overall energy consumption of BSs in RANs. Our previous work \cite{li_gm-pab:_2012} has shown the energy consumption of a BS is not linearly proportional to the traffic loads within its coverage area. Moreover, the energy consumption of BSs consists of two categories: some constant energy consumption stays irrelevant to BS's traffic loads while the remainder varies proportionally to BS's traffic loads. Hence, we adopt the generalized energy consumption model \cite{son_base_2011}, which can be summarized as
\begin{equation}
C_{\text{ee}}(\bm{\rho},\mathcal{B}_{\text{on}})=\sum_{i\in \mathcal{B}_{\text{on}}}\left[(1-q_i)\rho_iP_i+q_iP_i\right],
\label{eq:cost}
\end{equation}
where $\bm{\rho}=\{\rho_1,\cdots,\rho_N\}$. Besides, $q_i \in (0,1)$ is the constant power consumption percentage for BS $i$, and $P_i$ is the maximum power consumption of BS $i$ when it is fully utilized. 

On the other hand, in order to avoid the potential quality of service (QoS) deterioration, we introduce a delay-optimal metric in \cite{kim_alpha-optimal_2010} to demonstrate the flow performance.  As defined in \cite{kim_alpha-optimal_2010},  the delay-optimal performance function can be formulated as
\begin{equation}
\label{eq:delay_performance}
C_{\text{dp}} (\bm{\rho},\mathcal{B}_{\text{on}})=\sum_{i\in \mathcal{B}_{\text{on}}}\frac{\rho_i}{1-\rho_i}.
\end{equation}
Specifically, for a queue system $M/G/s$, \eqref{eq:delay_performance} equals the number of flows in the system. If we try to minimize \eqref{eq:delay_performance}, Little's law \cite{leon-garcia_probability_2008} implies that it's actually equivalent to minimize the average delay.

Above all, our problem is to find an optimal set of active BSs and corresponding user associations that minimizes the  function of the energy consumption while ensuring the QoS, namely
\begin{equation}
\label{eq:total_cost}
\begin{aligned}
&\min_{\bm{\rho} ,\mathcal{B}_{\text{on}}}\left\{C=
C_{\text{ee}}(\bm{\rho},\mathcal{B}_{\text{on}}) +\varsigma C_{\text{dp}}(\bm{\rho},\mathcal{B}_{\text{on}}) 
\right\},\\
&s.t.\,\,\rho_i \in [0,1) \,\, \forall i \in \mathcal{B},
\end{aligned}
\end{equation}
where $\varsigma$ is a positive balancing parameter with a unit W/s. $\varsigma$ indicates the equivalent cost for one flow waiting in the system and reflects the importance of the delay performance relative to the energy consumption.
\section{Stochastic BS switching operation in reinforcement learning framework}
\label{sec:learning}
\subsection{Markov decision process}
An MDP is defined as a tuple $M=<\mathcal{S},\mathcal{A},\mathcal{P},C>$, where $\mathcal{S}$ is the state space, $\mathcal{A}$ is the action space, $\mathcal{P}$ is a state transition probability function, and $C$ is a cost function. Specifically, at stage $k$, the traffic load state is $\bm{s}^{(k)}$. Following an action $\bm{a}^{(k)}=\{a^{(k)}_1, \cdots, a^{(k)}_N\}$, the controller choose to turn a BS $i\in \mathcal{B}$ into sleeping mode if  $a^{(k)}_i=0$. Otherwise, if $a^{(k)}_i=1$, the BS $i$ remains active. The users correspondingly associate themselves with the remaining active BSs $\mathcal{B}_{\text{on}}$ according to an indicator set $\bm{I}^{(k)}$, which can be determined by the specific metrics to select the serving BS, such as cell traffic loads or received signal strength, etc \cite{son_base_2011}. Thereafter, if we assume that as the traffic loads emerge, the traffic load state transforms into $\bm{s}^{(k+1)}$, which is determined by the exact volume of varying traffic loads at stage $k$ and the related serving BSs, with the transition probability
\begin{equation}
\label{eq:transitionProb}
\mathcal{P}(\bm{s}'|\bm{s}^{(k)},\bm{a}^{(k)}) =
\left\{ 
\begin{aligned}
1, & \ \bm{s}' = \bm{s}^{(k+1)};\\
0, & \ \text{otherwise}.\\
\end{aligned}
 \right.
\end{equation}
 Meanwhile, the immediate cost generated by the environment (computed by \eqref{eq:total_cost}) is fed back to the BS switching operation controller. 

The goal is to find a strategy $\pi$, which maps a state $\bm{s}$ to an action $\pi(\bm{s})$, i.e., $\bm{a}^{(k)}$, to minimize the discounted accumulative cost starting from the state $\bm{s}$. Formally, this accumulative cost is called as a state value function, which can be calculated by \cite{sutton_reinforcement_1998}
\begin{equation}
\label{eq:definition_state_value_function}
V^{\pi}(\bm{s}) =E_{\pi}\left[\sum\limits_{k=0}^{\infty} \gamma^k C(\bm{s}^{(k)},\pi(\bm{s}^{(k)})|\bm{s}^{(0)}=\bm{s})\right]=E_{\pi}\left[C(\bm{s},\pi(\bm{s}))+\gamma \sum_{\bm{s}' \in \mathcal{S}} \mathcal{P}(\bm{s}'|\bm{s},\pi(\bm{s}))V^{\pi}(\bm{s}')\right],
\end{equation}
where the positive parameter $\gamma$ is the discount factor that maps the future cost to the current state. Given the diminishing importance of future cost than the current one, $\gamma$ is smaller than 1. The optimal strategy $\pi^{\ast}$ satisfies the Bellman equation \cite{sutton_reinforcement_1998}:
\begin{equation}
V^{\ast}(\bm{s})=V^{\pi^{\ast}}(\bm{s})=\min_{\bm{a}\in\mathcal{A}}\left\{E_{\pi^\ast}\left[C(\bm{s},\bm{a})+\gamma \sum_{\bm{s}' \in \mathcal{S}} \mathcal{P}(\bm{s}'|\bm{s},\bm{a})V^{\pi^*}(\bm{s}')\right]\right\}.
\end{equation}
Since the optimal strategy minimizes the cumulative cost from the beginning, it contributes to design a foresighted energy saving scheme.

\subsection{The actor-critic learning framework for energy saving}
\begin{figure}
\centering
\includegraphics[width=0.75\textwidth]{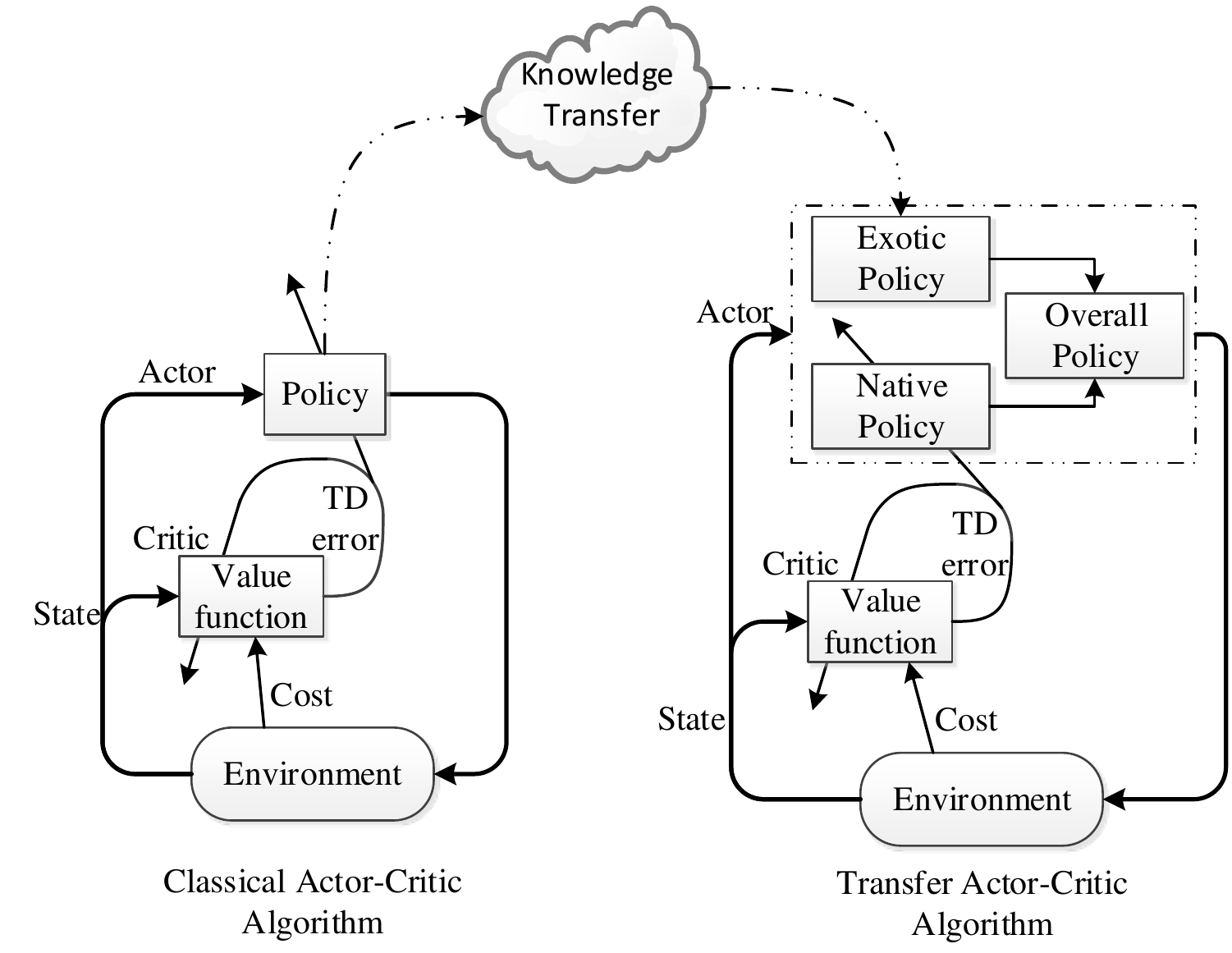}
\caption{Architecture of classical actor-critic algorithm and transfer actor-critic algorithm (TACT).}
\label{fig:learningArchitecture}
\end{figure}

There have been some well-known methods to solve the MDP issues such as dynamic programming \cite{sutton_reinforcement_1998}. Unfortunately, these methods heavily depend on prior knowledge of the environmental dynamics. However, it's challenging to know the future traffic loads precisely in advance. Therefore, in this paper, we employ reinforcement learning approaches to solve the MDP problem without requiring the knowledge of traffic loads a prior and specifically adopt the actor-critic algorithm. As the name implies, the actor-critic algorithm encompasses three components: actor, critic, and environment as illustrated in Fig. \ref{fig:learningArchitecture} (Left). At a given state, the actor selects an action in a stochastic way and then executes it. This execution transforms the state of environment to a new one with certain probability, and feeds back the cost to the actor. Then, the critic criticizes the action executed by the actor and updates the value function through a time difference (TD) error. After the criticism, the actor will update the policy to prefer the action with a smaller cost, and vice versa. The algorithm repeats the above procedure until convergence. The reasons to adopt actor-critic algorithm are three-folded: (i) since it generates the action directly from the stored policy, it requires little computation to select an action to perform; (ii) it can learn an explicitly stochastic policy which may be useful in non-Markov traffic variation environment of RANs \cite{zhou_robust_2011}; (iii) it separately updates the value function and policy \cite{sutton_reinforcement_1998}. As a result, it would be more easily to implement the policy knowledge transfer in Section \ref{sec:tact}, compared to other critic-only algorithms like $\varepsilon$-greed and Q-learning \cite{grondman_survey_2012}, .

\begin{figure*}
\centering
\includegraphics[width=0.95\textwidth]{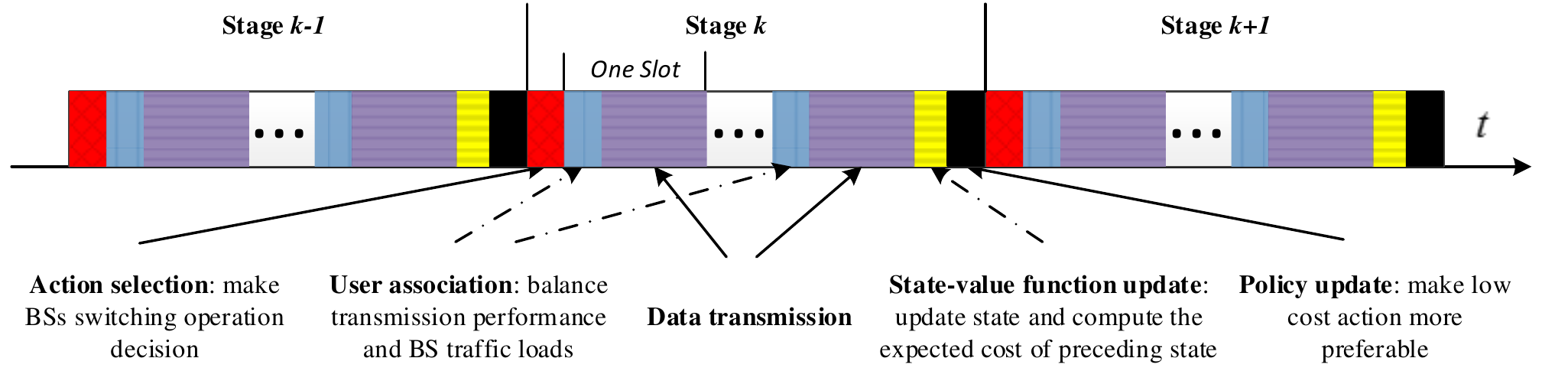}
\caption{Illustration of learning framework for energy saving.}
\label{fig:timeSlot}
\end{figure*}

We design an actor-critic learning framework for energy saving scheme as illustrated in Fig. \ref{fig:timeSlot}. 

(i) Action selection: Beforehand, let's assume that the system is at the beginning of stage $k$. Meanwhile, the traffic load state is $\bm{s}^{(k)}$. Thereafter, the controller needs to select an action according to a stochastic strategy, the purpose of which is to improve performance while explicitly balancing two competing objectives: a) searching for a better BS switching operation (exploration) and b) taking as little cost as possible (exploitation). As a result, the controller not only performs a good BS switching operation based on its past experience, but also is able to explore a new one. The most common methodology is to use a Boltzmann distribution. The controller chooses an action $\bm{a}$ in state $\bm{s}^{(k)}$ of stage $k$ with probability \cite{sutton_reinforcement_1998}
\begin{equation}
\label{eq:bolztmann}
\pi^{(k)}(\bm{s}^{(k)},\bm{a})=\frac{\exp\{p(\bm{s}^{(k)},\bm{a})/\tau\}}{\sum_{\bm{a}' \in \mathcal{A}}\exp\{p(\bm{s}^{(k)},\bm{a}')/\tau\}},
\end{equation}
where $\tau$ is a positive parameter called temperature. In addition, $p(\bm{s}^{(k)},\bm{a}^{(k)})$ indicates the tendency to select action $\bm{a}^{(k)}$ at the state $\bm{s}^{(k)}$, and it will update itself after every stage. It's worthwhile to note that though there exists the possibility that the remaining active BSs are not enough to serve the traffic loads in the present stage $k$. However, as the conventional energy saving scheme commonly does, the controller can start an emergent response paradigm to quickly turn on some BSs. Hence, in this paper, we assume the action $\bm{a}^{(k)}$, which the controller finally chooses, can meet the traffic load requirements.

(ii) User association and data transmission: In one stage, there exist several slots for user association and data transmission. After the controller chooses to turn some of BSs into sleeping mode and broadcasts the traffic load density at stage $k-1$, the users choose to connect one BS according to the modified metric in \cite{son_base_2011} and start the communications slot by slot.  Specifically, users at location $x$ choose to join BS $i^{\ast}$, while $i^{\ast}$ satisfies
\begin{equation}
\label{eq:user_association}
i^{\ast}(x)=\arg\max_{j\in \mathcal{B}_{\text{on}}} \frac{c_j(x,\mathcal{B}_{\text{on}})}{(1-q_j)P_j+\varsigma (1-\rho^{{k}-1}_j)^{-2}}, \,\, \forall x\in \mathcal{L}.
\end{equation}
 As stated in \cite{son_base_2011}, \eqref{eq:user_association} proves to be optimal to achieve the minimum of total cost in \eqref{eq:total_cost} if the active BSs are determined. Intuitively, \eqref{eq:user_association} would be simplified to $i^{\ast}(x)=\arg\max_{j\in \mathcal{B}_{\text{on}}} \frac{c_j(x,\mathcal{B}_{\text{on}})}{(1-q_j)P_j}, \,\, \forall x\in \mathcal{L}$, if we merely consider the minimization of energy consumption (i.e., $\varsigma=0$). The simplified equation implies that users at location $x$ prefer to choose to join the BS with the largest transmission rate at the same traffic load-related power consumption.

(iii) State-value function update: After the transmission part of stage $k$, the traffic loads in each BS will change, thus transforming the system to state $\bm{s}^{(k+1)}$ by \eqref{eq:transitionProb}. Meanwhile, the total cost for the transmission would be $C^{(k)}(\bm{s}^{(k)},\bm{a}^{(k)})$. Consequently, a TD error $\delta(\bm{s}^{(k)},\bm{a}^{(k)})$ would be computed by the difference between the state-value function $V^{(k)}(\bm{s}^{(k)})$ estimated at the preceding state and $C^{(k)}(\bm{s}^{(k)},\bm{a}^{(k)})+\gamma\cdot V^{(k)}(\bm{s}^{(k+1)})$ at the critic, namely
\begin{equation}
\label{eq:TDerror}
\begin{aligned}
&\delta^{(k)}(\bm{s}^{(k)},\bm{a}^{(k)})=C^{(k)}(\bm{s}^{(k)},\bm{a}^{(k)})+\gamma \sum_{\bm{s}' \in \mathcal{S}} \mathcal{P}(\bm{s}'|\bm{s}^{(k)},\bm{a}^{(k)})V(\bm{s}') - V(\bm{s}^{(k)}) \\
&=C^{(k)}(\bm{s}^{(k)},\bm{a}^{(k)})+\gamma\cdot V^{(k)}(\bm{s}^{{(k+1)}}) -V^{(k)}(\bm{s}^{(k)}).
\end{aligned}
\end{equation}
Afterwards, the TD error would feed back to the actor. By the way, the state-value function would be updated as
\begin{equation}
\label{eq:statevalueUpdate}
V^{(k+1)}(\bm{s}^{(k)})= V^{(k)}(\bm{s}^{(k)})+\alpha(\nu_1(\bm{s}^{(k)},k)) \cdot \delta^{(k)}(\bm{s}^{(k)},\bm{a}^{(k)}).
\end{equation}
Here, $\nu_1(\bm{s}^{(k)},k)$ denotes the occurrence times of state $\bm{s}^{(k)}$ in these $k$ stages. $\alpha(\cdot)$ is a positive step-size parameter that affects the convergence rate. On the other hand, if $\bm{s}\neq \bm{s}^{(k)}$, $V^{(k+1)}(\bm{s})$ will be kept the same as $V^{(k)}(\bm{s})$.

(iv) Policy update: At the end of stage $k$, the critic would employ the TD error to ``criticize'' the selected action, which is implemented as
\begin{equation}
\label{eq:policy_update}
p^{(k+1)}(\bm{s}^{(k)},\bm{a}^{(k)})=p^{(k)}(\bm{s}^{(k)},\bm{a}^{(k)})-\beta(\nu_2(\bm{s}^{(k)},\bm{a}^{(k)},k)) \cdot \delta^{(k)}(\bm{s}^{(k)},\bm{a}^{(k)}),
\end{equation} 
Similar to $\nu_1(\bm{s}^{(k)},k)$, $\nu_2(\bm{s}^{(k)},\bm{a}^{(k)},k)$ indicates the executed times of action $\bm{a}^{(k)}$ at state $\bm{s}^{(k)}$ in these $k$ stages. $\beta(\cdot)$ is a positive step-size parameter. \eqref{eq:bolztmann} and \eqref{eq:policy_update} ensure one action under a specific state can be selected with higher probability if the ``foresighted'' cost it takes is comparatively smaller, i.e., $\delta(\bm{s}^{(k)})<0$. Additionally,  if $\bm{a}\neq \bm{a}^{(k)}$,  $p^{(k+1)}(\bm{s}^{(k)},\bm{a})=p^{(k)}(\bm{s}^{(k)},\bm{a})$.

If each action is executed infinitely often in every state, in other words, if in the limit, the learning strategy is greedy with infinite exploration, the value function $V(\bm{s})$ and strategy $\pi^{(k)}(\bm{s},\bm{a})$ will finally converge to  $V^{\ast}$ and $\pi^{\ast}$ with probability (w.p.) 1 as $k \rightarrow \infty$ \cite{singh_convergence_2000}.

\section{Transfer actor-critic algorithm for stochastic BS switching operation }
\label{sec:tact}
\subsection{Motivation and formulation of transfer actor-critic algorithm}
The previous section addresses the methodology to exploit the classical AC algorithm to conduct the BS switching operation, culminating in an effective energy saving strategy in the end. In this section, we present the means that the controller utilizes the knowledge of learned strategies during historical periods or neighboring regions to be in the groove of finding the optimal BS switching operations.

Basically, the policy, say $p(\bm{s},\bm{a})$, which finally determines the strategy $\pi(\bm{s},\bm{a})$ in one learning task, indicates the tendency of action $\bm{a}$ to be chosen in state $\bm{s}$. When the learning process converges, the tendency to choose a specific action $\bm{a}$ in a specific state is comparatively larger than that of other actions. In other words, it means that if the BS switching operation is conducted based on one learned strategy, the energy saving in the whole system tends to be optimized in the long run. Hence, if the knowledge of this policy $p(\bm{s},\bm{a})$ is transferred to another task, e.g., the knowledge transferred from Period 1 (source task) to Period 2 (target task) within the same region of interest in Fig. \ref{fig:BSDeploymentMotivation}, the controller in the target task can make an attempt by taking the same action $\bm{a}$ when the traffic loads come into state $\bm{s}$. Compared to learning from the scratch, the controller might directly make the wisest choice at the very beginning. However, in spite of the similarities between the source task and the target task, there might still exist some differences. For example, the system might come into the same state in two different tasks, whereas the traffic loads in the source task (e.g., Period 1) might be usually higher than that in the target one (e.g., Period 2). Hence, instead of staying on the chosen action $\bm{a}$ in source task, the controller in target task can make a more aggressive choice to turn more BSs into sleeping mode, thus saving more energy consumption. Consequently, in this case, the transferred policy guides in a negative manner. To avoid this underlying problem, the transferred policy should have a decreasing impact on choosing a certain action, once the controller has attempted to choose this action and nurtured its own learning experience. 

Taking the above considerations into account, we propose a new policy update method, named Transferred Actor-CriTic algorithm (TACT) as Fig. \ref{fig:learningArchitecture}.  In the TACT algorithm, the overall policy (i.e., $p_{\text{o}}$)  to select an action is divided as a native one $p_{\text{n}}$ and an exotic one $p_{\text{e}}$. Without loss of generality, let's assume that at stage $k$, the traffic load state is $\bm{s}^{(k)}$ and the chosen action is $\bm{a}^{(k)}$. Accordingly, the overall policy  $p_{\text{o}}$ is updated as
\begin{equation}
\label{eq:tact_policy_update}
 p_{\text{o}}^{(k+1)}(\bm{s}^{(k)},\bm{a}^{(k)})=\left[(1-\zeta(\nu_2(\bm{s}^{(k)},\bm{a}^{(k)},k)))p_{\text{n}}^{(k+1)}(\bm{s}^{(k)},\bm{a}^{(k)})+\zeta(\nu_2(\bm{s}^{(k)},\bm{a}^{(k)},k))p_{\text{e}}(\bm{s}^{(k)},\bm{a}^{(k)})\right]_{-p_{\text{t}}}^{p_{\text{t}}},
\end{equation}
where $\left[x\right]_{a}^{b}$ with $b>a$, denotes the Euclidean projection of $x$ onto the interval [a,b], i.e., $\left[x\right]_{a}^{b}=a$ if $x<a$; $\left[x\right]_{a}^{b}=b$ if $x>b$; and $\left[x\right]_{a}^{b}=x$ if $a\leq x \leq b$. In this case, $a=-p_{\text{t}}$ and $b=p_{\text{t}}$, with $p_{\text{t}}>0$. Additionally,   $p_{\text{o}}^{(k+1)}(\bm{s}^{(k)},\bm{a})=p_{\text{o}}^{(k)}(\bm{s}^{(k)},\bm{a}), \forall \bm{a} \in \mathcal{A} \text{ but } \bm{a} \neq \bm{a}^{(k)}$. Besides that, $p_{\text{n}}(\bm{s},\bm{a})$ still updates itself according to the classical actor-critic algorithm, namely \eqref{eq:policy_update}. 

Initially, the exotic policy $p_{\text{e}}(\bm{s},\bm{a})$ dominates in the overall strategy. Hence, when the environment enters a state $\bm{s}$, the presence of $p_{\text{e}}(\bm{s},\bm{a})$ contributes to choose the action, which might be optimal to $\bm{s}$ in the source task. Consequently, the proposed policy update method leads to a possible performance jumpstart. On the other hand, since $\zeta(\cdot)\in (0,1)$ is the transfer rate and $\zeta(k)\rightarrow 0$ as $k\rightarrow \infty$, the effect of the transferred exotic policy $p_{\text{e}}(\bm{s},\bm{a})$ continuously decreases. Therefore, the controller can not only take advantage of the learned expertise in the source task, but also swiftly get rid of the negative guidelines. 

Finally, we summarize our proposed TACT algorithm in Algorithm \ref{tb:learning_framework} .

\begin{algorithm}                 
\caption{TACT : The Transfer Learning Framework for Energy Saving}          
\label{tb:learning_framework}                           
\begin{algorithmic}         
\STATE \textbf{Initialization}: 
\FOR{each $\bm{s} \in \mathcal{S}$, each $\bm{a} \in \mathcal{A}$}
\STATE Initialize state-value function $V(\bm{s})$, native policy function $p_{\text{n}}(\bm{s},\bm{a})$,  exotic policy function $p_{\text{e}}(\bm{s},\bm{a})$ (transferred knowledge) and strategy function $\pi(\bm{s},\bm{a})$;
\ENDFOR
\STATE \textbf{Repeat until convergent}
\begin{enumerate}
\item Choose an action $\bm{a}^{(k)}$ in state $\bm{s}^{(k)}$ according to $\pi^{(k)}(\bm{s}^{(k)},\bm{a}^{(k)})$ in \eqref{eq:bolztmann};
\item Users at location $x$ connect one BS $i$ by \eqref{eq:user_association} and then start data transmission;
\item If $\rho_i \leq 1, \forall i \in \mathcal{L}$, the chosen action is feasible. The cost function $C(\bm{s}^{(k)},\bm{a}^{(k)})$ is calculated by \eqref{eq:total_cost}; otherwise, an emergent response paradigm starts as the conventional scheme does.
\item Identify the traffic loads and accordingly update state $\bm{s}^{(k)}\rightarrow \bm{s}^{(k+1)}$ and compute the TD error by \eqref{eq:TDerror};
\item Update the state-value function \eqref{eq:statevalueUpdate} for $\bm{s}=\bm{s}^{(k)}$;
\item Update the native policy function and the overall policy function by \eqref{eq:policy_update}  and \eqref{eq:tact_policy_update} for $\bm{s}=\bm{s}^{(k)}, \bm{a}=\bm{a}^{(k)}$, respectively;
\item Update the strategy function $\pi^{(k+1)}(\bm{s}^{(k)},\bm{a})$ by \eqref{eq:bolztmann}, for $\bm{s}=\bm{s}^{(k)}$ and all $\bm{a} \in \mathcal{A}$.
\end{enumerate}
\end{algorithmic}
\end{algorithm}

\subsection{Convergence analysis}
Next, we are interested in the convergence of TACT algorithm, since the knowledge transfer makes the policy update in the proposed TACT algorithm distinct from that in the classical AC algorithms and it becomes difficult to directly apply the convergence results in the latter ones.  We start the analysis by introducing several related lemmas.
Singh \cite{singh_convergence_2000} shows that the Boltzmann method is greedy in the limit with infinite exploration, based on a large enough $\tau$. Therefore, we have the following lemma.
\begin{lemma}
\label{lemma:glie}
If we use the Boltzmann exploration method with a large enough $\tau$, there thereby exists an $\eta>0$, such that
\begin{equation}
\lim\limits_{k\rightarrow \infty}\frac{\nu_2(\bm{s},\bm{a},k)}{k}\geq \eta, \forall \bm{s} \in \mathcal{S}, \bm{a} \in \mathcal{A}. 
\end{equation}
In other words, as $k\rightarrow \infty$, $\nu_2(\bm{s},\bm{a},k) =\eta k\rightarrow \infty$.
\end{lemma}
\begin{definition}
Define a function $\vartheta_{\bm{s},\bm{a}}(p_{\text{o}})$ as 
\begin{equation}
\vartheta_{\bm{s},\bm{a}}(p_{\text{o}}) =\left\{
\begin{aligned}
0 &\ \text{if}\  p_{\text{o}}(\bm{s},\bm{a})=p_{\text{t}} &\text{and} &\ \delta(\bm{s},\bm{a})\geq 0,\\
\  &\ \text{or}\   p_{\text{o}}(\bm{s},\bm{a})=-p_{\text{t}} &\text{and} &\ \delta(\bm{s},\bm{a})\leq 0,\\
1 &\ \text{otherwise}.
\end{aligned}
\right.
\end{equation}
\end{definition}

The next theorem states that our proposed policy update tracks an ordinary differential equation (ODE).
\begin{theorem}
\label{theorem:ode}
Assume that the learning rate $\beta(k)$ in \eqref{eq:policy_update} satisfies \begin{equation}
\label{eq:beta_assumption}
\sum_{k=0}^{\infty}\beta(k) =\infty, \beta(k)\geq 0, \sum_{k=0}^{\infty}\beta(k)^2 < \infty , 
\end{equation}
and the transfer rate $\zeta(k)$ satisfies $\lim \zeta(k)/\beta(k)\rightarrow 0$ as $k\rightarrow \infty$. $p_{\text{o}}(\bm{s},\bm{a})$ asymptotically tracks the solution of the ODE
\begin{equation}
\label{eq:ode}
\dot{p}_{\text{o}}(t)=-\delta(\bm{s},\bm{a})\vartheta_{\bm{s},\bm{a}}(p_{\text{o}}), \forall \bm{s} \in \mathcal{S}, \bm{a} \in \mathcal{A},
\end{equation} 
where $\delta(\bm{s},\bm{a})=\lim\delta^k(\bm{s},\bm{a})$ as $k\rightarrow \infty$.
\end{theorem}
\begin{proof}
We provide a proof sketch here and will address the details in Appendix. Without loss of generality, assume that the state is $\bm{s}^{(k)}$. By Algorithm \ref{tb:learning_framework}, at stage $k$, the policy value $p_{\text{o}} (\bm{s}^{(k)},\bm{a})$ would be changed only when $\bm{a}$ is the executed action $\bm{a}^{(k)}$. Therefore, by merely including the updated values, we could form another discrete sequence $\hat{p}_{\text{o}}(\hat{k})$\footnote{Indeed, $\hat{p}_{\text{o}}(\hat{k})$ refers to $\hat{p}_{\text{o}}^{(\hat{k})}(\bm{s}^{(k)},\bm{a}^{(k)})$. But, for simplicity of representation, the notation of $\bm{s}^{(k)}$ and $\bm{a}^{(k)}$ is omitted here.} to indicate the evolution of $p_{\text{o}}^{(k)} (\bm{s}^{(k)},\bm{a}^{(k)})$. Here, the index $\hat{k}$ equals $\nu_2(\bm{s}^{(k)},\bm{a}^{(k)},k)$. After that, by introducing a concept of $\beta(\hat{k})$-induced continuous time $t$ and interpolating the discrete sequence $\hat{p}_{\text{o}}(\hat{k})$,  we construct a continuous sequence $\hat{p}^{(0)}(t)$ and its shifted version $\hat{p}^{(\hat{k})}(t)$. Next, we prove that the shifted continuous sequence $\hat{p}^{(\hat{k})}(t)$ is equicontinuous. Based on the discussions around the Arzel\`a-Ascoli Theorem \cite{kushner_stochastic_2003}, we finally obtain that any limit of $\hat{p}(t)$, or the discrete equivalent $\hat{p}_{\text{o}}(\hat{k})$,  must track the solution of the ODE in \eqref{eq:ode} for a sufficiently large $\hat{k}$. By Lemma \ref{lemma:glie}, the theorem comes.
\end{proof}
In addition, we introduce the definition of a \textit{strict Lyapunov function}  \cite{kushner_stochastic_2003}, which is the fundamental of our following proof.
\begin{definition}
\label{df:lyapunov}
Suppose that for an ODE $\dot{z}(t)=f(z)$ defined on a region $\mathcal{D}$, $V(z)$ is a continuously differentiable and real-valued function of $z$ such that $V(0)=0, V(z)>0, \forall z\neq 0$. If $\dot{V}(t)=\nabla V \cdot \dot{z}(t)=\nabla V \cdot f(z)\leq 0$ on the region $\mathcal{D}$, and the equality holds only when $\dot{z}(t)=0$, the function $V(z)$ is a strict Lyapunov function for the ODE $\dot{z}(t)$.
\end{definition}
Our proof relies on the following theorem by Konda and Borkar \cite{konda_actor-critic--type_1999}, which establishes the convergence of a general actor-critic algorithm.
\begin{theorem}
\label{theorem:convergence}
Assume that the learning rate $\alpha(k)$ satisfies the assumptions in Section 2.2 \cite{konda_actor-critic--type_1999} and $\beta(k)$ and $\zeta(k)$ meet the conditions in Theorem  \ref{theorem:ode}. If the strategy $\pi$, which is derived by \eqref{eq:bolztmann} with the policy update method given by \eqref{eq:tact_policy_update}, has a \textit{strict Lyapunov function} for the ODE $\dot{\pi}(t)$, we thereby have $\pi$ converges w.p. 1 and $\left\Vert \pi -\pi^{\ast}\right\Vert \leq \epsilon$ for any $\epsilon >0$ as $p_{\text{t}}\rightarrow \infty$. 
\end{theorem}

Beforehand, it comes the following lemma by directly applying \eqref{eq:definition_state_value_function} in \eqref{eq:TDerror}.
\begin{lemma}
\label{lemma:sum_equal_zero}
\begin{equation}
\label{eq:sum_equal_zero}
\sum\limits_{\bm{a}\in\mathcal{A}}\delta(\bm{s},\bm{a})\pi(\bm{s},\bm{a})=0, \forall \bm{s}\in \mathcal{S}.
\end{equation}
\end{lemma}

\begin{lemma}
\label{lemma:leq_equation_ode_lyap}
If the strategy $\pi(\bm{s},\bm{a})$ tracks the solution of ODE $\dot{\pi}(t)$, and $\dot{\pi}(t)$ satisfies $\dot{\pi}(t)\delta(\bm{s},\bm{a}) \leq 0$, then we have $\nabla V^{\pi}(\bm{s})  \dot{\pi}(t)  \leq \dot{\pi}(t)\delta(\bm{s},\bm{a}) \leq 0, \forall \bm{s}\in \mathcal{S}$.
\end{lemma}
\begin{proof}
For two distinct policies $\pi$ and $\pi'$, let's define a value function operation $T(\pi',V^{\pi}(\bm{s}))$\\
$=E_{\pi'}\left[C(\bm{s},\bm{a})+\gamma \sum_{\bm{s}' \in \mathcal{S}} p(\bm{s}'|\bm{s},\bm{a})V^{\pi}(\bm{s}')\right]$. Assume that there exists an infinitesimal $\epsilon>0$ such that $\pi+\epsilon \dot{\pi}(t)$ is still a valid strategy. If denote $\pi'=\pi+\epsilon \dot{\pi}(t)$, we thereby have
\begin{displaymath}
\begin{aligned}
&T(\pi',V^{\pi}(\bm{s}))-V^{\pi}(\bm{s})=E_{\pi'}\left[C(\bm{s},\bm{a})+\gamma \sum_{\bm{s}' \in \mathcal{S}}\mathcal{P}(\bm{s}'|\bm{s},\bm{a})V^{\pi}(\bm{s}')\right]-V^{\pi}(\bm{s})\\
&=\sum\limits_{\bm{a}\in\mathcal{A}} \left\{\pi'\left[C(\bm{s},\bm{a})+\gamma \sum_{\bm{s}' \in \mathcal{S}}\mathcal{P}(\bm{s}'|\bm{s},\bm{a})V^{\pi}(\bm{s}')-V^{\pi}(\bm{s})\right]\right\}\\
&=\sum\limits_{\bm{a}\in\mathcal{A}}  (\pi+\epsilon \dot{\pi}(t))\delta(\bm{s},\bm{a})\\
&=\sum\limits_{\bm{a}\in\mathcal{A}}  \epsilon \dot{\pi}(t)\delta(\bm{s},\bm{a})\leq 0\\
\end{aligned}
\end{displaymath}
The last equality follows from Lemma \ref{lemma:sum_equal_zero}.

Denote an iteration operation of $T(\pi',V^{\pi}(\bm{s}))$ as $T^{n}(\pi',V^{\pi}(\bm{s}))=T^{n-1}(\pi',T(\pi',V^{\pi}(\bm{s})))$, we have $T^n (\pi',V^{\pi}(\bm{s})) \leq T^{n-1} (\pi',V^{\pi}(\bm{s})) \leq \cdots \leq V^{\pi}(\bm{s})$.

 In addition, $T^n (\pi',V^{\pi}(\bm{s})) - V^{\pi}(\bm{s}) \leq \sum\limits_{\bm{a}\in\mathcal{A}}  \epsilon \dot{\pi}(t)\delta(\bm{s},\bm{a})$, for $n > 1$. As $n\rightarrow \infty$, $T^n(\pi',V^{\pi}(\bm{s}))$ $\rightarrow$ $V^{\pi'}(\bm{s})$, we obtain
\begin{displaymath}
\frac{V^{\pi'}(\bm{s})-V^{\pi}(\bm{s})}{\epsilon}=\frac{V^{\pi+\epsilon \dot{\pi}}(\bm{s})-V^{\pi}(\bm{s})}{\epsilon}\leq \dot{\pi}(t)\delta(\bm{s},\bm{a})\leq 0.
\end{displaymath}
As $\epsilon \rightarrow 0$, $\nabla V^{\pi}(\bm{s})  \dot{\pi}(t) \leq \dot{\pi}(t)\delta(\bm{s},\bm{a}) \leq 0.$ The claim follows.
\end{proof}

\begin{theorem}
\label{theorem:lyap_function}
$\sum\limits_{\bm{s}\in\mathcal{S}}V^{\pi}(\bm{s})$ is a strict Lyapunov function for ODE $\dot{\pi}(t)$, if $p_{\text{t}}$ is sufficiently large.
\end{theorem}
\begin{proof}
By explicit differentiating \eqref{eq:bolztmann} over $t$, we have
\begin{displaymath}
\begin{aligned}
&\dot{\pi}(t)=\frac{\frac{1}{\tau}\exp\left[p_{\text{o}}(\bm{s},\bm{a})/\tau \right]}{\sum_{\bm{a}'\in \mathcal{A}} \exp\left[p_{\text{o}}(\bm{s},\bm{a}')/\tau \right]} \dot{p}_{\text{o}}(t) - \frac{\frac{1}{\tau}\exp\left[p_{\text{o}}(\bm{s},\bm{a})/\tau \right] \sum_{\bm{a}'\in \mathcal{A}} \left\{\exp\left[p_{\text{o}}(\bm{s},\bm{a}')/\tau \right] \dot{p}_{\text{o}}(t)\right\} }{\left\{\sum_{\bm{a}'\in \mathcal{A}} \exp\left[p_{\text{o}}(\bm{s},\bm{a}')/\tau \right]\right\}^2}  \\
&=\frac{1}{\tau} \pi(\bm{s},\bm{a}) \dot{p}_{\text{o}}(t) -\frac{1}{\tau}  \pi(\bm{s},\bm{a})\sum_{\bm{a}'\in \mathcal{A}} \pi(\bm{s},\bm{a}') \dot{p}_{\text{o}}(t) \\
&=\frac{1}{\tau} \pi(\bm{s},\bm{a}) \dot{p}_{\text{o}}(t) -\frac{1}{\tau}  \pi(\bm{s},\bm{a})\sum_{\bm{a}'\in \mathcal{A}} \pi(\bm{s},\bm{a}') \left[-\delta(\bm{s},\bm{a}')\vartheta_{\bm{s},\bm{a}'}(p_{\text{o}})\right]\\
&=\frac{1}{\tau} \pi(\bm{s},\bm{a}) \dot{p}_{\text{o}}(t) .
\end{aligned}
\end{displaymath}
The last equality follows from Lemma \ref{lemma:sum_equal_zero}, after taking into account that if $p_{\text{t}}$ is sufficiently large, $\vartheta_{\bm{s},\bm{a}}(p_{\text{o}})=1$ holds.
By Theorem \ref{theorem:ode}, 
\begin{displaymath}
\begin{aligned}
\dot{\pi}(t)\delta(\bm{s},\bm{a}) &= \left[-\frac{1}{\tau} \pi(\bm{s},\bm{a}) \delta(\bm{s},\bm{a})\vartheta_{\bm{s},\bm{a}}(p_{\text{o}})\right]\cdot \delta(\bm{s},\bm{a})\\
&=-\frac{1}{\tau} \pi(\bm{s},\bm{a})\left[\delta(\bm{s},\bm{a})\right]^2\leq 0.
\end{aligned}
\end{displaymath}
The equality only holds at the equilibrium point $\dot{p}_{\text{o}}(t)=-\delta(\bm{s},\bm{a})=0$. By Lemma \ref{lemma:leq_equation_ode_lyap}, $ \nabla V^{\pi}(\bm{s})  \dot{\pi}(t)  \leq 0$. Therefore, according to Definition \ref{df:lyapunov}, the claim follows.
\end{proof}

\begin{theorem}
Regardless of any initial value chosen for $p_{\text{n}}(\bm{s},\bm{a})$, and transferred knowledge $p_{\text{e}}(\bm{s},\bm{a})$, if the learning rate $\alpha(k)$, $\beta(k)$ and the transfer rate $\zeta(k)$ meets the required conditions meanwhile $p_{\text{t}}$ and $\tau$ are sufficiently large, the Algorithm \ref{tb:learning_framework} converges.
\end{theorem}
\begin{proof}
The proof is the direct application of Theorem \ref{theorem:convergence}, which establishes the convergence given two conditions. First, the policy $p_{\text{o}}(\bm{s},\bm{a})$ tracks the solution of an ODE, by Theorem \ref{theorem:ode}. Second, the tracked ODE has a strict Lyapunov function, by Theorem \ref{theorem:lyap_function}. Therefore, the learning process in Algorithm \ref{tb:learning_framework} converges.
\end{proof}
\section{Numerical analysis}
\label{sec:analysis}
\begin{figure}
\centering
\includegraphics[width=0.75\textwidth]{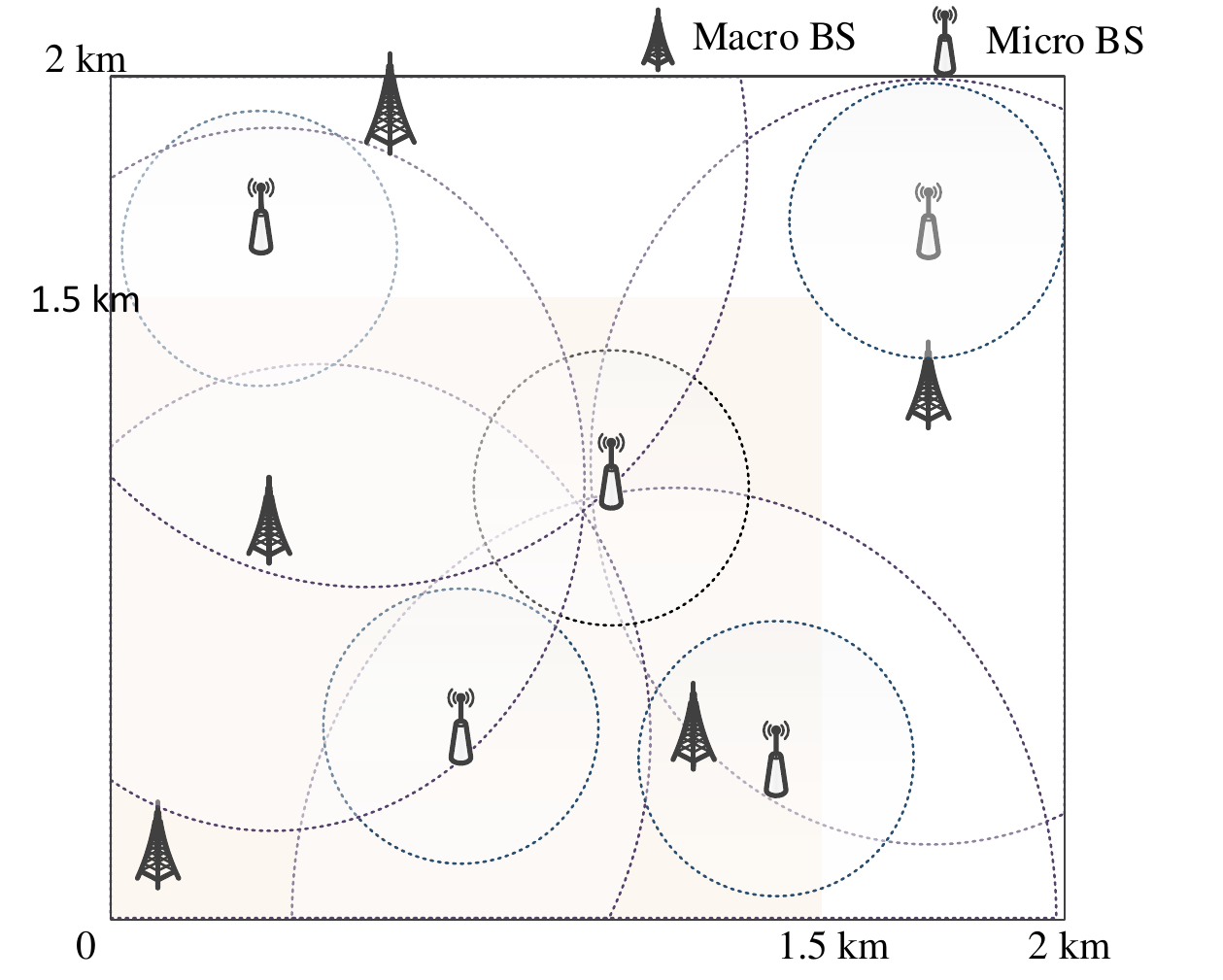}
\caption{Illustration of BS deployment in our simulation scenario.}
\label{fig:BSdeployment}
\end{figure}
We validate the energy efficiency improvement of our proposed scheme by extensive simulations under practical configurations. Here, we simulate for an area of 2km $\times$ 2km, where there exist five macro BSs and five micro BSs \cite{son_speedbalance:_2012,son_base_2011} as Fig. \ref{fig:BSdeployment} shows. Moreover, we assume that file transmission requests at location $x \in \mathcal{L}$ follow a Poisson point process with arrival rate $\lambda(x)$ and file size $1/\mu(x)$. To ease the simulation process, each BS' traffic load state merely takes value of 0 or 1 (0 represents the case where the realistic traffic loads are smaller than the historical average one while 1 indicates the other cases). Beyond that, we assume the maximum transmission powers for BSs, i.e., 20W and 1W for macro and micro BSs, respectively. Based on the linear relationship between transmission and operational energy consumption in \cite{son_base_2011}, the maximum operational powers for macro BS and micro BS are 865W and 38W, respectively. We set the propagation channel according to the COST-231 modified Hata model \cite{ieee_802.16_boradband_wireless_access_working_group_ieee_2008} and don't consider the influence of fast fading effect and noise. As for the proposed TACT algorithm, the learning rate $\alpha(k)=1/k$ while $\beta(k)=1\Big/(k\log k)$ \cite{konda_actor-critic--type_1999}. Moreover, the transfer rate $\zeta(k)=\theta^k$, with the transfer rate factor $\theta \in (0,1)$, thus satisfying the assumption in Theorem \ref{theorem:ode}. 


\begin{table}
\centering
\begin{threeparttable}
\caption{Used simulation parameters} 
\label{tb:para}
\begin{tabular}{@{}rl@{}}
\toprule
\textbf{Parameter description} & \textbf{Value}\\
\midrule
Simulation area & 2km $\times$ 2km  \\
Maximum transmission power & \tiny Macro BS  \footnotesize 20W \tiny Micro BS \footnotesize 1W\\
Maximum operational power & \tiny Macro BS  \footnotesize 865W \tiny Micro BS \footnotesize 38W\\ 
Height & \tiny Macro BS  \footnotesize 32m \tiny Micro BS \footnotesize 12.5m\\  
Channel bandwidth & 1.25MHz\\
Intra-cell interference factor & 0.01 \\
\midrule
Arrival rate $\lambda(x)$ & $5 \times 10^{-6}$ \\
File size $1/\mu(x)$ &  $100$kbyte\\
Constant Power Percentage $q$  & 0.5\\
Temperature $\tau$ & 1000 \\
Discount Factor $\gamma$ & 0.001\\
Transfer Rate $\theta$ & 0.2\\
Delay Performance Importance $\zeta$ & 0 W/s\\
\bottomrule
\end{tabular}
\end{threeparttable}
\end{table}

By the way, we assume the extra cost is negligible when we turn the necessary BSs into active mode. Besides, we define \textit{cumulative energy consumption ratio} (CECR) as the metric to test how much energy saving can be achieved due to the application of our proposed schemes. Specifically, the CECR metic is defined by the ratio of the accumulative energy consumption when certain BSs are turned off (as our scheme runs) to that when all the BSs stay active since our simulation starts. This definition is reasonable since it can show the foresighted energy efficiency improvement, which is exactly the goal of an energy saving scheme.

Besides, we would compare the performance\footnote{Due to the space limitation, only temporal knowledge transfer is considered for the TACT scheme.} of our proposed schemes with that of the state-of-the-art (SOTA) scheme \cite{son_base_2011}, which assumes the controller can obtain a full knowledge of traffic loads in prior and finds the optimal BS switching solution by greedily turning as many as BSs into sleeping mode. To simplify the comparison, we simulate by adjusting only one parameter while configuring the others according to Table \ref{tb:para}.

\begin{figure}
\centering
\includegraphics[width=0.75\textwidth]{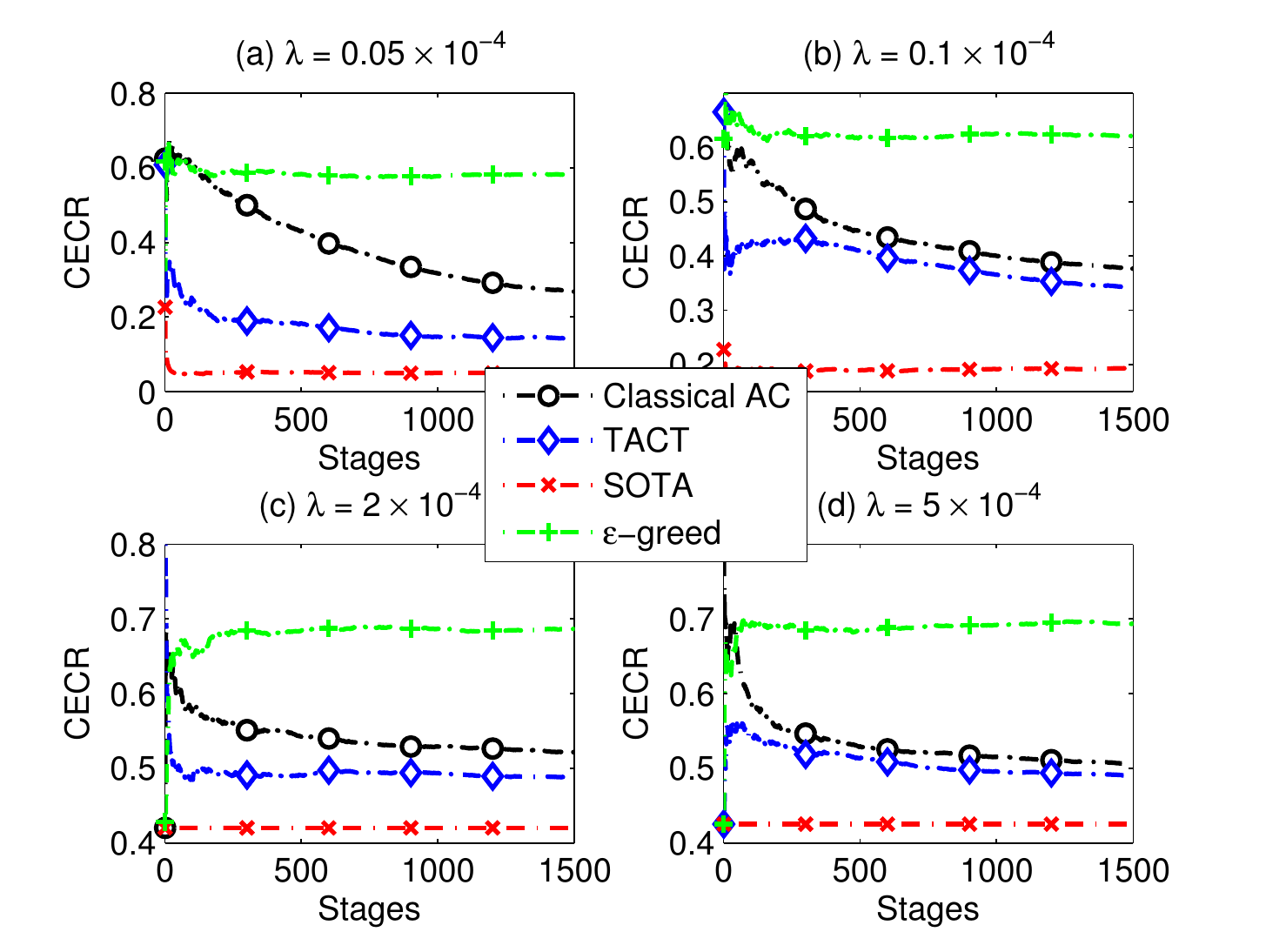}
\caption{Performance comparison under various homogeneous traffic arrival rates.}
\label{fig:staticCom}
\end{figure}

Firstly, we examine how much energy saving can be achieved under different static traffic load arrival rates. \cite{son_base_2011} shows that when all BSs are turned on, a homogeneous traffic distribution of $\lambda(x)=10^{-4}$ for all $x \in \mathcal{L}$ will offer loads corresponding to about 10\% of BSs utilizations. Therefore, we vary the homogeneous traffic arrival rate $\lambda(x)$ from $5 \times 10^{-6}$ to $5 \times 10^{-4}$ to reflect the effect of traffic loads on energy saving. Here, the transferred policy is generated from a source task with the static arrival rate $\lambda=5\times 10^{-6}$. As depicted in Fig. \ref{fig:staticCom}, we can expect more significant energy conservation with the decrease of arrival rate $\lambda$. This is because that if all the BSs stay active under lower traffic loads, the BSs are more highly under-utilized. Moreover, the CECR continues decreasing as the simulation runs, since the controller will have a better understanding of the traffic loads and thereby know whichever action has a better energy efficiency. Unfortunately, since the proposed learning schemes\footnote{Together with the classical AC algorithm, $\varepsilon$-Greed algorithm is also compared. But, due to the insufficient exploration issue in the $\varepsilon$-Greed algorithm \cite{sutton_reinforcement_1998}, the corresponding performance is the worst in all cases and then refrains us to use it further.} are performed without the knowledge of traffic loads a prior, the performance of them are inferior to that of the SOTA scheme, especially at the beginning of the simulations. However, we can see that the gap compensated for the absent knowledge becomes much smaller, when the TACT scheme is applied with the learned knowledge. 

\begin{figure}
\centering
\includegraphics[width=0.75\textwidth]{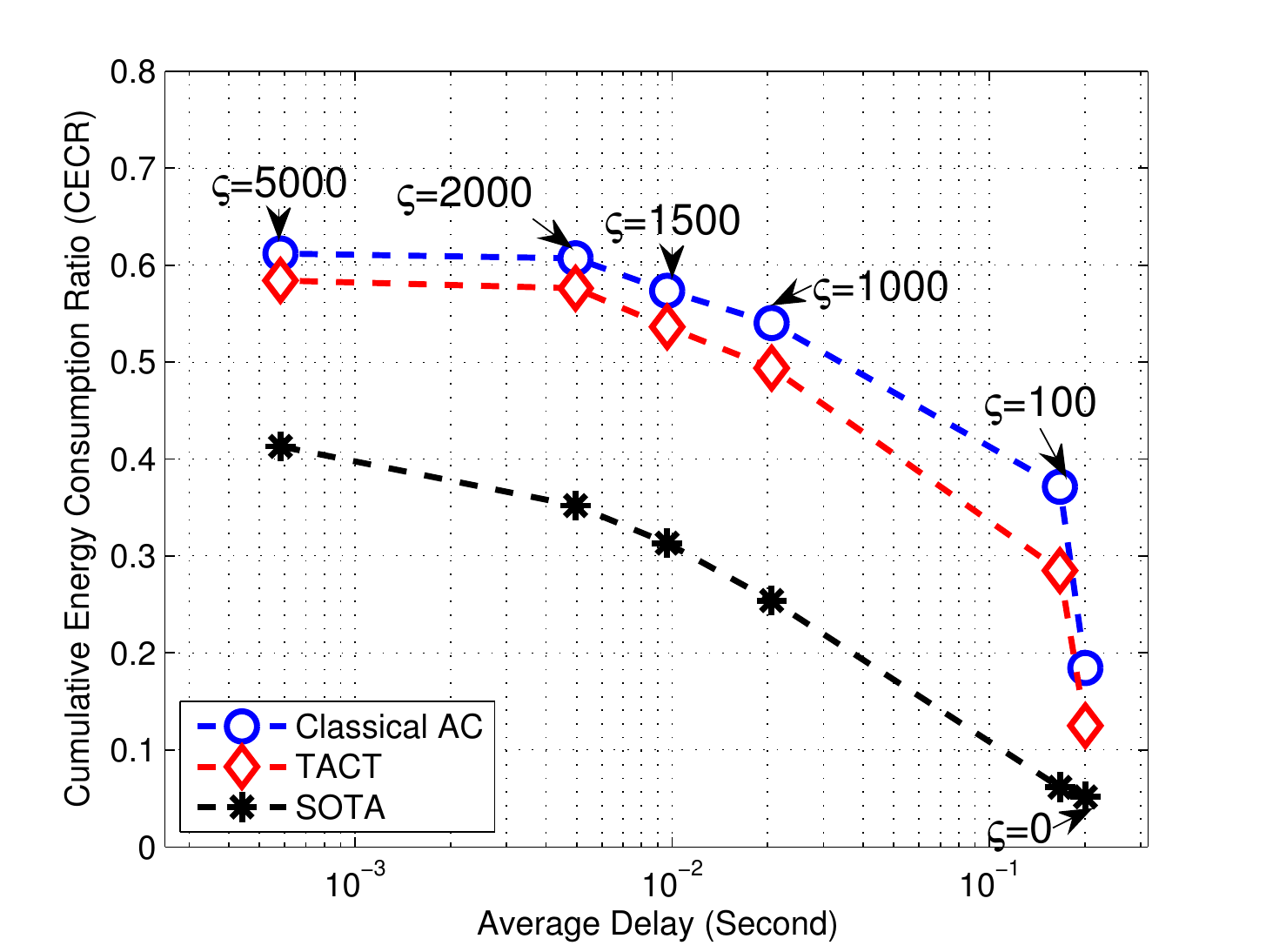}
\caption{Performance tradeoff between energy and delay under different delay equivalent cost scenarios.}
\label{fig:performanceVarsigma}
\end{figure}

After validating the feasibility of proposed learning framework to save the energy, Fig. \ref{fig:performanceVarsigma} depicts the performance tradeoff between energy consumption and delay under different delay equivalent cost scenarios by tuning $\varsigma$. When $\varsigma=0$, the energy saving is most significant. However, this also incurs a limited increase in delay. Comparatively, the energy saving would be less obvious if we put more emphasis on the delay equivalent cost by choosing a larger $\varsigma$ so as to decrease the delay. Again, we could also find that the tradeoff points of TACT are closer to those of the SOTA solutions in all these scenarios. 

\begin{figure}
\centering
\includegraphics[width=0.75\textwidth]{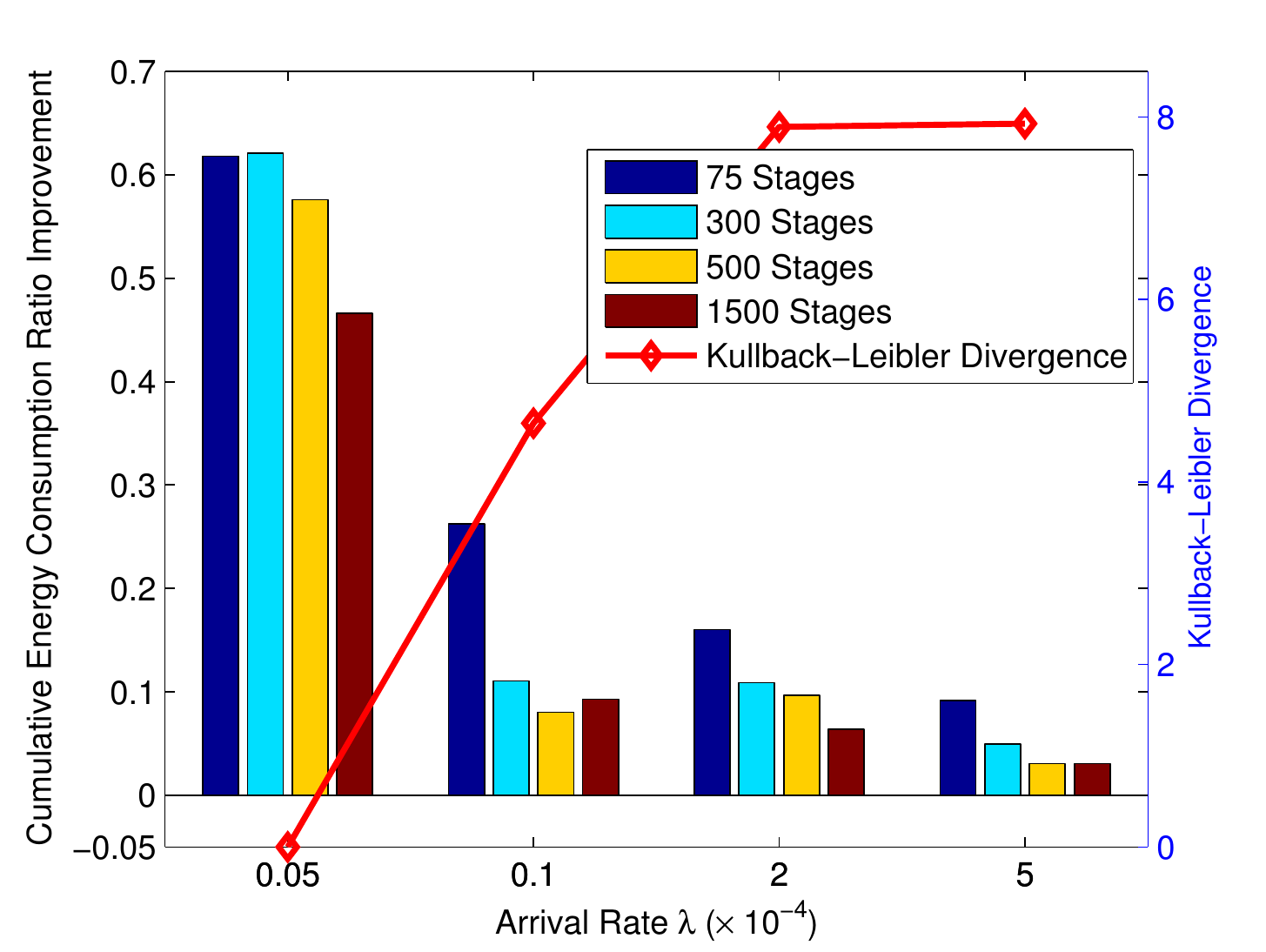}
\caption{Performance improvement of TACT scheme over classical AC scheme versus Kullback-Leibler divergence. The bars corresponding to the left Y-axis reflect the CECR improvement while the dotted line corresponding the right Y-axis represents the Kullback-Leibler Divergence.}
\label{fig:KLDiverrgence}
\end{figure}

\begin{figure}
\centering
\includegraphics[width=0.75\textwidth]{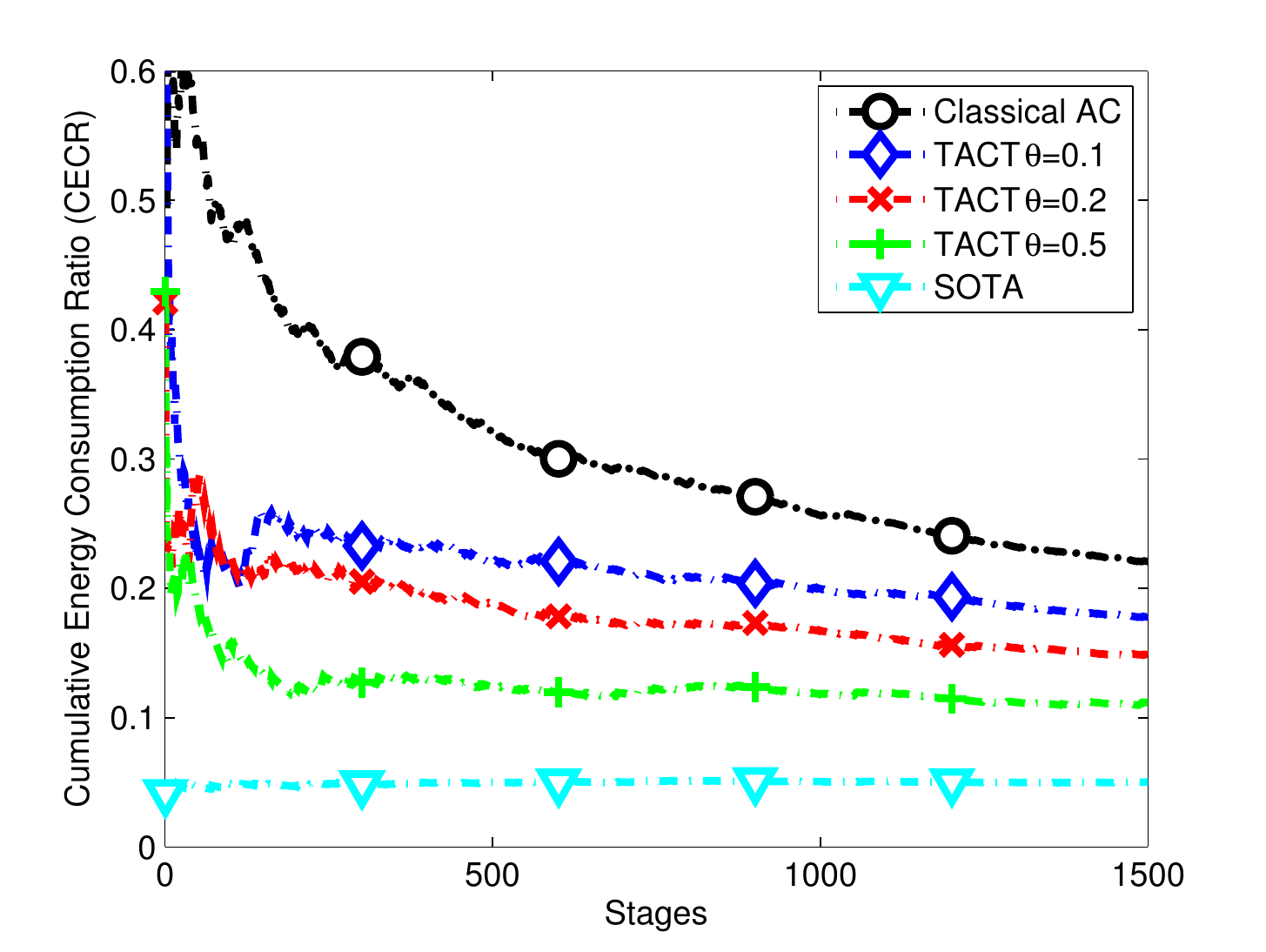}
\caption{Performance impact of the transfer rate factor $\theta$ to the TACT scheme.}
\label{fig:performanceTheta}
\end{figure}

Fig. \ref{fig:KLDiverrgence} presents the performance improvement\footnote{The performance improvement is calculated by dividing the energy consumption margin between TACT scheme and classical AC scheme over the energy consumption using classical AC scheme.} of TACT scheme over classical AC scheme. As expected, the TACT scheme yields a relatively large performance improvement, especially at the beginning of each simulation. In other words, the TACT scheme contributes to a performance jumpstart, or a faster convergence speed. Fig. \ref{fig:KLDiverrgence} also depicts the similarity between the source task and the target task, measured by Kullback-Leibler divergence \cite{kullback_information_1951}. It shows a smaller Kullback-Leibler divergence between the source task and the target task leads to a more efficient transfer effect. Besides, we also plot the impact of transfer rate factor $\theta$ in Fig. \ref{fig:performanceTheta}. Generally speaking, as we expect, larger $\theta$ results in faster convergence rate and larger energy saving.  

\begin{figure}
\centering
\includegraphics[width=0.75\textwidth]{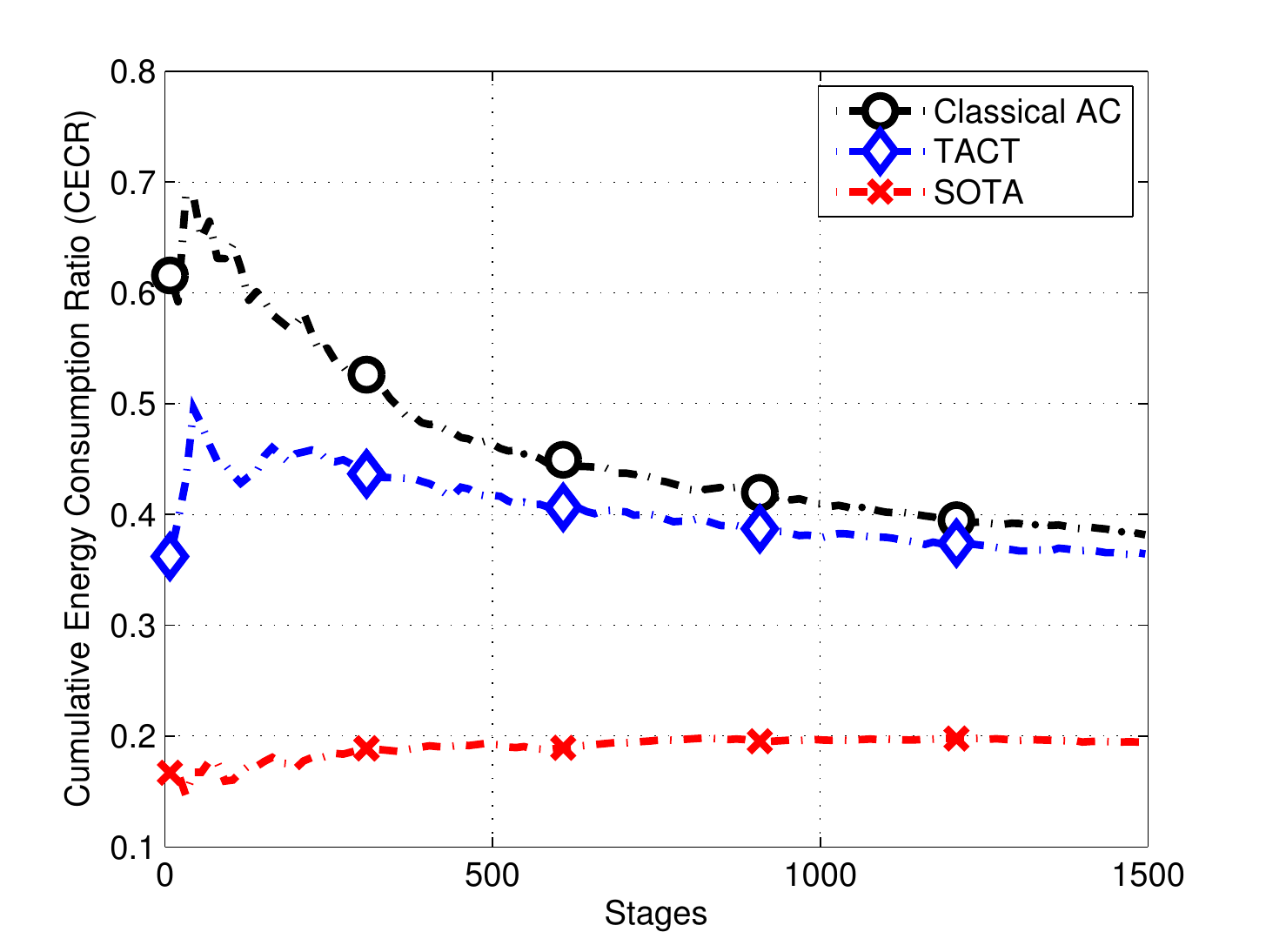}
\caption{Performance comparison with time-variant traffic arrival rate $\overline{\lambda}(k,x)=(0.99\cdot\cos(2\pi(k+10)/24)+1)\times 10^{-5}$.}
\label{fig:variantCom}
\end{figure}

We also investigate the performance of the proposed schemes when traffic loads periodically fluctuates. \cite{oh_energy_2010} shows practical traffic load profile is periodical and can be approximated by a sinusoidal function $\overline{\lambda}(k)=\lambda_\text{V} \cdot \cos(2\pi(k+\phi)/D)+\lambda_\text{M}$, where $D$ is the period of a traffic load profile, $\lambda_\text{V}$ is the variance of traffic profile and $\lambda_\text{M}$ is the mean arrival rate. Therefore, we employ $\overline{\lambda}(k,x)=(0.99\cdot\cos(2\pi(k+10)/24)+1)\times 10^{-5}$ to approximate the practical traffic load arrival rate at location $x\in \mathcal{L}$. Fig. \ref{fig:variantCom} compares the performance of the proposed schemes and shows that the TACT scheme converges faster than the classical AC scheme.

\begin{figure}
\centering
\includegraphics[width=0.8\textwidth]{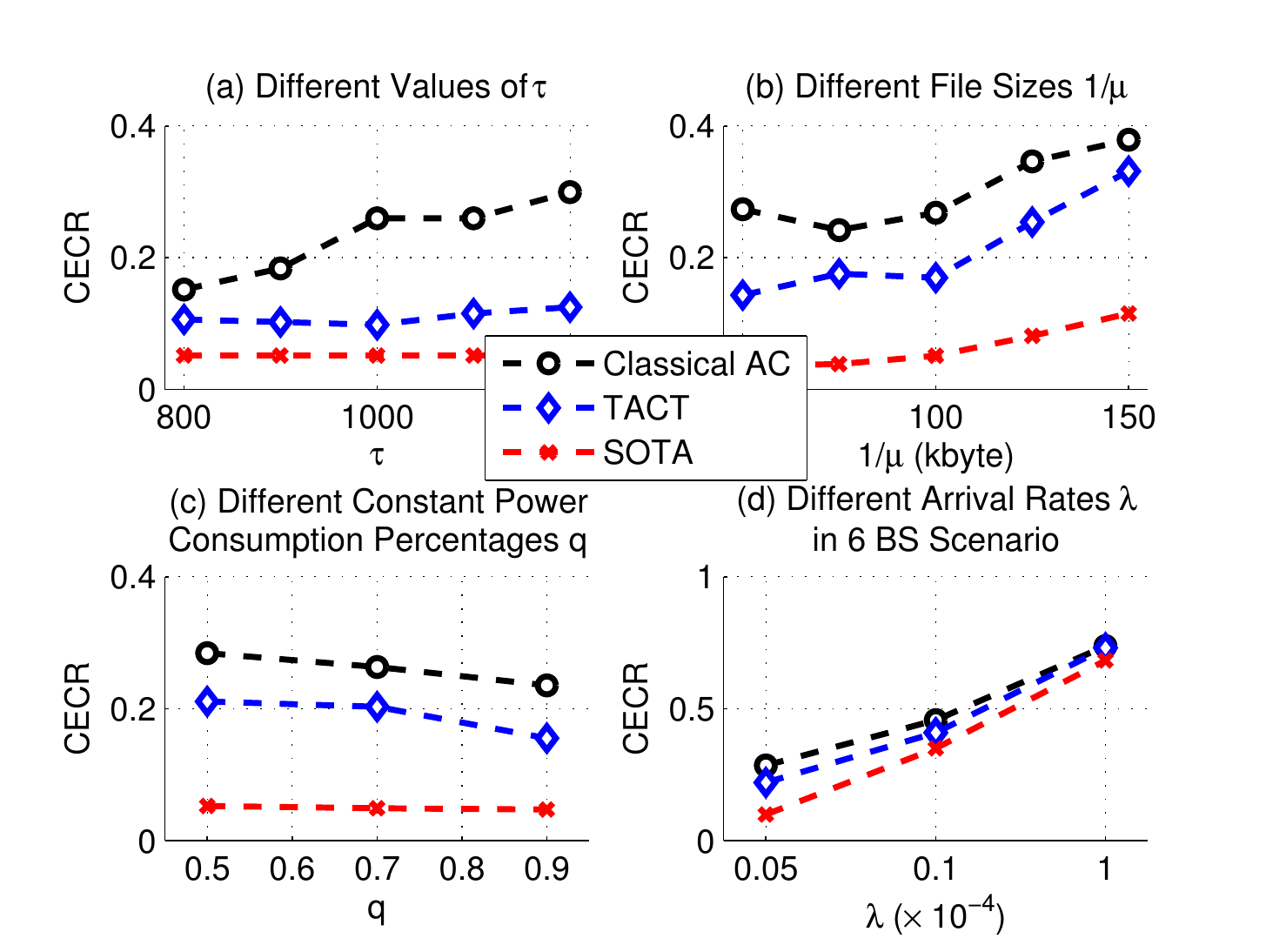}
\caption{Performance comparison under various configurations: (a) Different values of $\tau$, (b) Different file sizes $1/\mu$, (c) Different energy consumption models, and (d) Different arrival rates in a 6 BS (red shaded region in Fig. \ref{fig:BSdeployment}) scenario. All these simulations results are generated after 1500 stages.}
\label{fig:sensitivityAnalysis}
\end{figure}

At last, we continue the performance evaluation of our proposed schemes and present more detailed sensitivity analyses in Fig. \ref{fig:sensitivityAnalysis}. In Fig. \ref{fig:sensitivityAnalysis} (a)--(c), we present the simulation results under various configurations to reflect the effect of temperature value $\tau$, file size $1/\mu$ and constant power consumption percentage $q$. We can observe that the performance trends match our common sense in all these cases. For example, in Fig. \ref{fig:sensitivityAnalysis} (a), a larger value of $\tau$ implies that the controller has a higher desire to explore new actions. Therefore, even though the controller has tried the wisest action, the controller would choose more actions with larger cost, resulting in a less significant energy consumption saving. Fortunately, the TACT scheme could exploit the transferred knowledge to avoid some certainly undesirable actions and performs better than the classical AC one, especially at larger values of $\tau$. Fig. \ref{fig:sensitivityAnalysis} (b) demonstrates that the effect of file sizes to the scheme performance would be similar to that of arrival rates. Fig. \ref{fig:sensitivityAnalysis} (c) implies that the schemes will perform better when the constant power consumption accounts for a larger proportion of the whole cost, since turning off one under-utilized BS will make a clearer difference and save more energy in these cases. On the other hand, we also give the simulation results for the red shaded region with 6 BSs (illustrated in Fig. \ref{fig:BSdeployment}) and exhibit the robustness of our proposed schemes in different BS deployment scenarios.

\section{Conclusion}
\label{sec:conclusion}
In this paper, we have developed a learning framework for BS energy saving. We specifically formulated the BS switching operations under varying traffic loads as a Markov decision process. Besides, we adopt the actor-critic method, a reinforcement learning algorithm, to give the BS switching solution to decrease the overall energy consumption. Afterwards, to fully exploit the temporal relevancy in traffic loads, we propose a transfer actor-critic algorithm to improve the strategies by taking advantage of learned knowledge from historical periods. Our proposed algorithm provably converges given certain restrictions that arise during the learning process, and the extensive simulation results manifest the effectiveness and robustness of our energy saving schemes under various practical configurations.

Similar to the simulated  temporal knowledge transfer, our proposed TACT approach is potentially viable to be applied in spatial scenarios to achieve a performance improvement. Unfortunately, the mapping of knowledge will be sometimes less straightforward in the latter case, due to the underlying BS geographical deployment differences. Therefore, we are dedicated to handle the related meaningful yet more challenging issues over spatial knowledge transfer in the future. 

\section*{Acknowledgment}
The authors would like to express their sincere gratitude to the editor Prof. Jack L. Burbank and the anonymous reviewers for their kind comments. The authors also thank Qianlan Ying (ZJU), Shun Cai (SEU) and Yi Zhong (USTC) for their commendable suggestions in improving the paper quality. This paper is supported by the National Basic Research Program of China (973Green, No. 2012CB316000), the Key (Key grant) Project of Chinese Ministry of Education (No. 313053), the Key Technologies R\&D Program of China (No. 2012BAH75F01), and the grant of ``Investing for the Future'' program of France ANR to the CominLabs excellence laboratory (ANR-10-LABX-07-01).

\section*{Appendix}
Proof of Theorem \ref{theorem:ode}.
\begin{proof}
Without loss of generality, assume that at stage $k$, the state is $\bm{s}^{(k)}$ and the chosen action is $\bm{a}^{(k)}$. Moreover, the latest stage that the state-action pair $(\bm{s}^{(k)},\bm{a}^{(k)})$ occurred is stage $m$. Thus, by Algorithm \ref{tb:learning_framework}, the policy $p_{\text{o}}^{(j)}(\bm{s}^{(k)},\bm{a}^{(k)})$ remains invariant for any $j\in[m,\cdots,k)$. For simplicity of representation, we denote one sequence $\hat{p}_{\text{o}}({\hat{k}})=p_{\text{o}}^{(k)}(\bm{s}^{(k)},\bm{a}^{(k)})$ and $\hat{p}_{\text{o}}({\hat{k}-1})=p_{\text{o}}^{(j)}(\bm{s}^{(k)},\bm{a}^{(k)})$ for any $j\in[m,\cdots,k)$, where the index $\hat{k}$ equals $\nu_2(\bm{s}^{(k)},\bm{a}^{(k)},k)$.  In addition, the sequences $\hat{p}_{\text{n}}({\hat{k}})$ and $\hat{\delta}({\hat{k}})$ are defined analogously to $\hat{p}_{\text{o}}({\hat{k}})$. Thus, based on \eqref{eq:tact_policy_update}, we have
\begin{equation}
\label{eq:policy_update_previous_stage}
\begin{aligned}
\hat{p}_{\text{o}}(\hat{k})&=p_{\text{o}}^{(k)}(\bm{s}^{(k)},\bm{a}^{(k)})\\
&=\left[(1-\zeta(\hat{k}-1))\hat{p}_{\text{n}}({\hat{k}})+\zeta(\hat{k}-1)p_{\text{e}}(\bm{s}^{(k)},\bm{a}^{(k)})\right]_{-p_{\text{t}}}^{p_{\text{t}}}.\\
\end{aligned}
\end{equation}

Firstly, assume that $p_{\text{t}}$ is large enough such that $\left\vert \hat{p}_{\text{o}}(\hat{k}) \right\vert<p_{\text{t}}$ and  $\left\vert \hat{p}_{\text{o}}(\hat{k}+1)\right\vert<p_{\text{t}}$, while the assumption will be dropped later. 

Subtracting \eqref{eq:tact_policy_update} to \eqref{eq:policy_update_previous_stage}, we obtain
\begin{equation}
\label{eq:differential_discrete}
\begin{aligned}
&\hat{p}_{\text{o}}(\hat{k}+1) -\hat{p}_{\text{o}}(\hat{k})\\
&=(1-\zeta(\hat{k}-1))\left(
\hat{p}_{\text{n}}(\hat{k}+1)-\hat{p}_{\text{n}}(\hat{k})\right) -(\zeta(\hat{k})-\zeta(\hat{k}-1))\left(\hat{p}_{\text{n}}(\hat{k}+1)-p_{\text{e}}(\bm{s}^{(k)},\bm{a}^{(k)})\right)\\
&=-\beta(\hat{k})(1-\zeta(\hat{k}-1))\hat{\delta}(\hat{k})-(\zeta(\hat{k})-\zeta(\hat{k}-1))\left(\hat{p}_{\text{n}}(\hat{k}+1)-p_{\text{e}}(\bm{s}^{(k)},\bm{a}^{(k)})\right).\\
\end{aligned}
\end{equation}
The last equality holds because of \eqref{eq:policy_update}.

\begin{figure}
\centering
\includegraphics[scale=0.7]{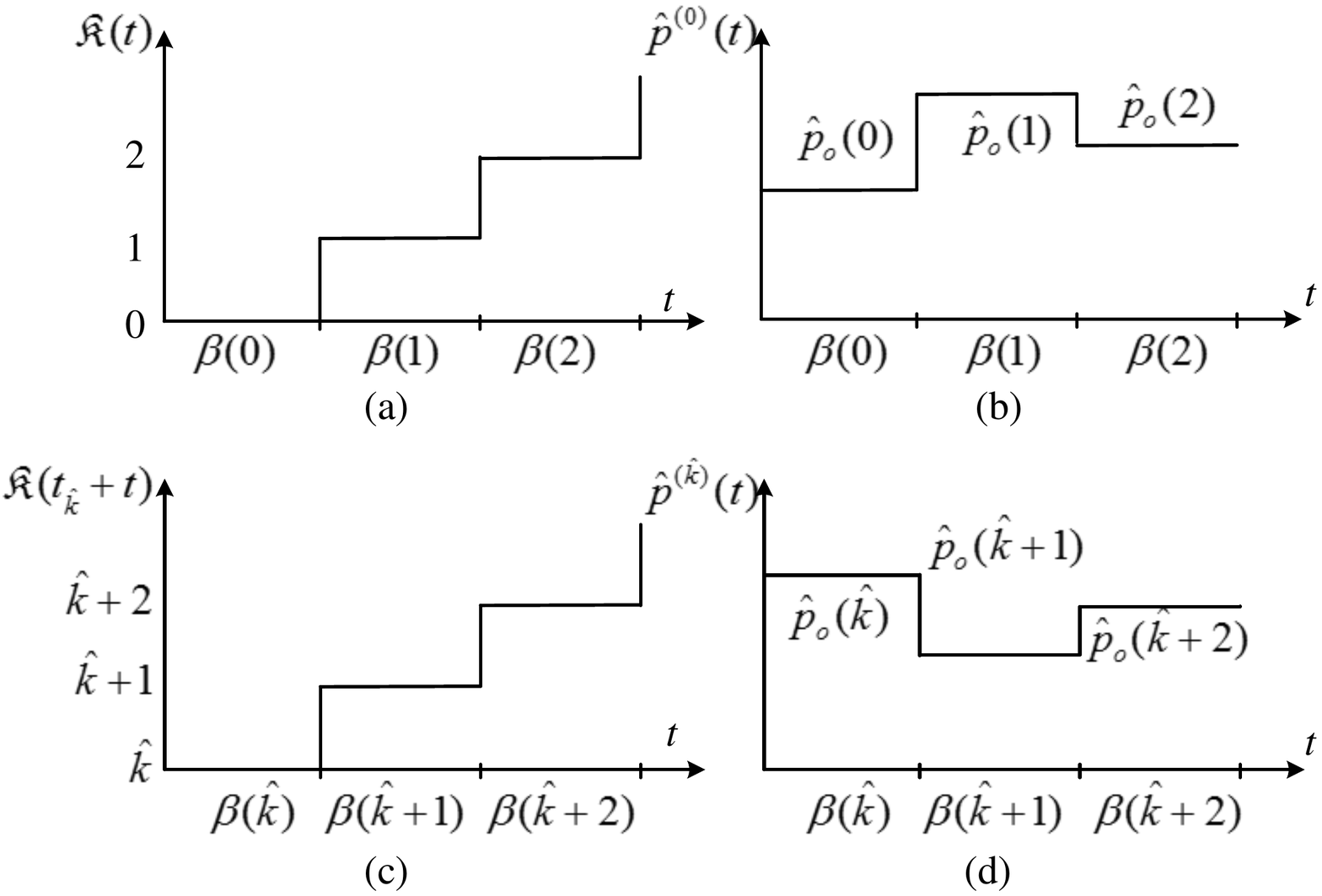}
\caption{Illustration of (a) the function $\mathfrak{K}(t)$, (b) the function $\hat{p}^{(0)}(t)$, (c) the function $\mathfrak{K}(t_{\hat{k}}+t)$ and (d) the function $\hat{p}^{(\hat{k})}(t)$.}
\label{fig:interpolation}
\end{figure}

Define $t_0=0$ and $t_{\hat{k}}=\sum_{j=0}^{\hat{k}-1}\beta(j)$. For $t\geq 0$, let $\mathfrak{K}(t)$ denote the unique value of $\hat{k}$ such that $t_{\hat{k}}\leq t < t_{\hat{k}+1}$, as Fig. \ref{fig:interpolation}-(a) depicts. For $t<0$, set $\mathfrak{K}(t)=0$. Define the \textit{continuous time interpolation} $\hat{p}^{(0)}(\cdot)$ on $(-\infty, \infty)$ by $\hat{p}^{(0)}(t)=p_{\text{o}}^{(0)}(\bm{s}^{(k)},\bm{a}^{(k)}) $ for $t\leq 0$, and for $t\geq 0$, 
\begin{displaymath}
\hat{p}^{(0)}(t)=\hat{p}_{\text{o}}(\mathfrak{K}(t))=\hat{p}_{\text{o}}(\hat{k}), \text{ for } t_{\hat{k}}\leq t < t_{\hat{k}+1}.
\end{displaymath}
Moreover, we define the \textit{sequence of shifted processes} $\hat{p}^{(\hat{k})}(t)=\hat{p}^{(0)}(t_{\hat{k}}+t), \  t\in (-\infty, \infty) $, as Fig. \ref{fig:interpolation}-(d) depicts. Define $Y_j=0$ and $Z_j=0$ for $j<1$. Moreover, define $Y_j=(1-\zeta(j-1))\hat{\delta}(j)$ and $Z_j=(\zeta(j)-\zeta(j-1))\left(\hat{p}_{\text{n}}(j+1)-p_{\text{e}}(\bm{s}^{(k)},\bm{a}^{(k)})\right)$ for $j\geq 1$. Define $Z^{(0)}(t)=0$ for $t\leq 0$ and 
\begin{displaymath}
\begin{aligned}
&Z^{(0)}(t)=\sum\nolimits_{j=0}^{\mathfrak{K}(t)-1}Z_j, \\ &Z^{(\hat{k})}(t)=Z^{(0)}(t_{\hat{k}}+t)-Z^{(0)}(t_{\hat{k}})=\sum\nolimits_{j=\hat{k}
}^{\mathfrak{K}(t_{\hat{k}}+t)-1}Z_j, \ t\geq 0.\\
\end{aligned}
\end{displaymath}

Taking into account the definitions above (recall that $\mathfrak{K}(t_{\hat{k}})=\hat{k}$), the following equation can be achieved by a manipulation of \eqref{eq:differential_discrete}
\begin{equation}
\label{eq:piecewise_p}
\begin{aligned}
&\hat{p}^{(\hat{k})}(t)=\hat{p}_{\text{o}}({\hat{k}})-\sum\nolimits_{j=\hat{k}
}^{\mathfrak{K}(t_{\hat{k}}+t)-1}\left(\beta(j)Y_j+Z_j \right)\\
&= \hat{p}_{\text{o}}({\hat{k}})-\sum\nolimits_{j=\hat{k}
}^{\mathfrak{K}(t_{\hat{k}}+t)-1}\left(\beta(j)Y_j\right)-Z^{(\hat{k})}(t).
\end{aligned}
\end{equation}
Since  $\hat{p}^{(\hat{k})}(t)$ is piecewise constant, we can rewrite \eqref{eq:piecewise_p} as
\begin{equation}
\hat{p}^{\hat{k}}(t)=\hat{p}_{\text{o}}({\hat{k}})-\int_{0}^{t} Y_{\mathfrak{K}(t_{\hat{k}}+x)}dx -Z^{(\hat{k})}(t)+\varphi^{(\hat{k})}(t),
\end{equation}
where $\varphi^{(\hat{k})}(t)$ is the outcome due to the replacement of the first sum in \eqref{eq:piecewise_p} by an integral. $\varphi^{(\hat{k})}(t)=0$ at the times when the interpolated sequences have jumps, i.e., $t=t_{\hat{k'}}-t_{\hat{k}},\ \hat{k'}>\hat{k}$,  and $\varphi^{(\hat{k})}(t)\rightarrow 0$ in $t$ as $\hat{k}\rightarrow \infty$ under the assumption in \eqref{eq:beta_assumption}.  

Besides that, by our assumption that $\lim \zeta(\hat{k})/\beta(\hat{k}) \rightarrow 0$ as $\hat{k}\rightarrow \infty$,  $Z_{\hat{k}}=(\zeta(\hat{k})-\zeta(\hat{k}-1))\cdot\left(\hat{p}_{\text{n}}(\hat{k}+1)-p_{\text{e}}(\bm{s}^{(k)},\bm{a}^{(k)})\right) =o(\beta(\hat{k}))\left(\hat{p}_{\text{n}}(\hat{k}+1)-p_{\text{e}}(\bm{s}^{(k)},\bm{a}^{(k)})\right)$. Therefore, $Z^{(\hat{k})}(t)=\sum\nolimits_{j=\hat{k}}^{\mathfrak{K}(t_{\hat{k}}+t)-1}o(\beta(j))\left(\hat{p}_{\text{n}}(j+1)-p_{\text{e}}(\bm{s}^{(k)},\bm{a}^{(k)})\right)$. Thus, as $\hat{k}\rightarrow \infty$, $Z^{(\hat{k})}(t)$ is negligible, since it's a small order of magnitude to $\sum\nolimits_{j=\hat{k}
}^{\mathfrak{K}(t_{\hat{k}}+t)-1}\beta(j)Y_j$.

Given the above discussion, as $\hat{k}\rightarrow \infty$, the sequence of functions $\hat{p}^{(\hat{k})}(t)=\hat{p}_{\text{o}}(\hat{k})-\int_{0}^{t} Y_{\mathfrak{K}(t_{\hat{k}}+x)}dx$ is equicontinous. Hence, by the Arzel\`a-Ascoli Theorem \cite{kushner_stochastic_2003}, there is a convergent subsequence in the sense of uniform convergence on each bounded time integral, and it's easily seen that any limit of $\hat{p}(t)$, or the discrete equivalent $\hat{p}_{\text{o}}(\hat{k})$,  must track the solution of the ODE $\dot{\hat{p}}(t)=-\hat{\delta}(\hat{k})$ for sufficiently large $\hat{k}$.

Next, in the special case where  $\hat{p}_{\text{o}}(\hat{k}-1)=p_{\text{t}}$ and $\hat{\delta}(\hat{k}-1)\geq 0$, at next stage $k$, the overall policy  $\hat{p}_{\text{o}}(\hat{k})$ would equal $p_{\text{t}}$. Thus, the ODE  $\dot{\hat{p}}(t)=0$. Similar discussion can be easily applied to the case, where $\hat{p}_{\text{o}}(\hat{k}-1)=-p_{\text{t}}$ and $\hat{\delta}(\hat{k}-1)\leq 0$. 

Furthermore, as $k\rightarrow 0$, by Lemma \ref{lemma:glie}, $\hat{k}=\nu_2(\bm{s}^{(k)},\bm{a}^{(k)},k)\rightarrow \infty$.

Summarizing the above discussion and taking into account $\delta(\bm{s}^{(k)},\bm{a}^{(k)})=\lim\delta^{(k)}(\bm{s}^{(k)},\bm{a}^{(k)})$ as $k \rightarrow \infty$, we can obtain
\begin{equation}
\dot{p}_{\text{o}}(t)=-\delta(\bm{s}^{(k)},\bm{a}^{(k)})\vartheta_{\bm{s}^{(k)},\bm{a}^{(k)}}(p_{\text{o}}).
\end{equation}
The claim follows.
\end{proof}
\bibliographystyle{IEEEtran}
\bibliography{IEEEabrv,draft}

\end{document}